\DeclareSymbolFont{rsfscript}{OMS}{rsfs}{m}{n}
\DeclareSymbolFontAlphabet{\mathrsfs}{rsfscript}
\renewcommand{\mathcal}{\mathrsfs}

\newcommand{\Ht}{\hat H}
\newcommand{\pf}{p^{\text{fix}}}
\newcommand{\pfl}{p_\ell^{\text{fix}}}
\newcommand{\un}{u^{(n)}}

\newcommand{\cna}[1][n]{\overline{C}^{(#1)}} 
\newcommand{\us}{w}

\newcommand{\LO}{\mathcal{L}}

\newcommand{\G}{\mathcal{\gamma}}
\newcommand{\nth}{$n^{\text{\tiny th}}$ }
\newcommand{\cel}{c_{\ell}}
\newcommand{\Deltad}{\Delta^{(d)}}
\newcommand{\Deltas}{\Delta^{(s)}}

\newcommand{\celn}{c_{\ell,0}}
\newcommand{\celcn}[1][n-1]{\overline{c_\ell C^{(#1)}}}

\newcommand{\ov}{b}

\documentclass[pre]
              {revtex4}

\usepackage{bm,color,nicefrac}
\usepackage[tight]{subfigure}
\usepackage[pdftex]{graphics}
\usepackage{epsfig}
\usepackage{amssymb}
\usepackage{amsmath}
\usepackage{bm}

\usepackage{xcolor}
\definecolor{TangoChameleon3}{HTML}{4E9A06}
\definecolor{TangoSkyBlue2}{HTML}{3465A4}
\definecolor{TangoScarletRed2}{HTML}{CC0000}
\definecolor{TangoAluminium2}{HTML}{D3D7CF}

\begin{document}

\title {Collective Fluctuations in models of adaptation}

\author{Oskar Hallatschek} \email{ohallats@berkeley.edu}
\affiliation{Biophysics and Evolutionary Dynamics Group, Departments
  of Physics and Integrative Biology, University of California,
  Berkeley, USA } \date{\today}
\author{Lukas Geyrhofer} 
\affiliation{Biophysics and Evolutionary Dynamics Group, Max Planck
  Institute for Dynamics and Self-Organization, G{\"o}ttingen, Germany}

\begin{abstract}
  The dynamics of adaptation is difficult to predict because it is
  highly stochastic even in large populations. The uncertainty emerges
  from number fluctuations, called genetic drift, arising in the small
  number of particularly fit individuals of the population. Random
  genetic drift in this evolutionary vanguard also limits the speed of
  adaptation, which diverges in deterministic models that ignore these
  chance effects. Several approaches have been developed to analyze
  the crucial role of noise on the \emph{expected} dynamics of
  adaptation, including the mean fitness of the entire population, or
  the fate of newly arising beneficial deleterious mutations. However,
  very little is known about how genetic drift causes fluctuations to
  emerge on the population level, including fitness distribution
  variations and speed variations. Yet, these phenomena control the
  replicability of experimental evolution experiments and are key to a
  truly predictive understanding of evolutionary processes. Here, we
  develop an exact approach to these emergent fluctuations by a
  combination of computational and analytical methods.  We show,
  analytically, that the infinite hierarchy of moment equations can be
  closed at any arbitrary order by a suitable choice of a dynamical
  constraint. This constraint regulates (rather than fixes) the
  population size, accounting for resource limitations. The resulting
  linear equations, which can be accurately solved numerically,
  exhibit fluctuation-induced terms that amplify short-distance
  correlations and suppress long-distance ones. Importantly, by
  accounting for the dynamics of sub-populations, we provide a
  systematic route to key population genetic quantities, such as
  fixation probabilities and decay rates of the genetic diversity. We
  demonstrate that, for some key quantities, asymptotic formulae can
  be derived. While it is natural to consider the process of
  adaptation as a branching random walk (in fitness space) subject to
  a constraint (due to finite resources), we show that other noisy
  traveling waves likewise fall into this class of constrained
  branching random walks. Our methods, therefore, provide a systematic
  approach towards analyzing fluctuations in a wide range of
  population biological processes, such as adaptation, genetic
  meltdown, species invasions or epidemics.
\end{abstract}

\maketitle

Many important evolutionary and ecological processes
rely on the behavior of a small number of individuals that have a
large dynamical influence on the population as a whole. This is,
perhaps, most obvious in the case of biological adaptation: Future
generations descend from a small number of currently well-adapted
individuals. The genetic footprint of the large majority of the
population is wiped out over time by the fixation of more fit
genotypes. These dynamics can be visualized as a traveling wave in
fitness space, see Fig.~\ref{Fig:noisy-wave-schema}A. At any time, the
currently most fit ``pioneer'' individuals reside in the small tip of
the wave. As time elapses, the wave moves towards higher fitness and
the formerly rare most fit individuals dominate the population. By
that time, however, a new wave tip of even more fit mutants has formed
and the cycle of transient dominance continues.

The principle of ``a few guiding the way for many'' also characterizes the
motion of flocks of birds, which can be controlled by just a few
leaders, or the expansion of an invasive species, which depends on
pioneers most advanced into the virgin territory. The overall dynamics
of these processes can become highly erratic even in large populations
because the behavior of the entire population is influenced by strong
number fluctuations, called genetic drift, occurring in the small
subset of ``pioneer'' individuals.

Such propagation processes with an extreme sensitivity of noise have
also been called ``pulled'' waves, because they are pulled along by
the action of the most advanced individuals~\footnote{In ``pushed''
  waves, by contrast, most of the growth occurs behind the front at
  higher population densities. While these ``pushed'' waves allow for
  simple mean-field approximation that neglect noise, ``pulled'' waves
  break down when noise is neglected. The reason is that noise is a
  singular perturbation and neglecting it can lead to qualitatively
  wrong predictions or even divergences.}. If one ignores the
fluctuations at the population level and is interested only in the
expected dynamics of the population, one might be tempted to simply
ignore genetic drift in models of pulled waves. However, it turns out
that mean-field models ignoring genetic drift drastically
overestimate the speed of traveling waves, to the point that they
predict an ever accelerating rather than a finite speed of
adaptation. It took 70 years since the first formulation of traveling
wave models by Fisher and Kolomogorov, to realize that genetic drift
influences both the expectation and the variation in singular
ways~\cite{tsimring1996rna,brunet1997shift}.

The \emph{expected} behavior of pulled waves has since been analyzed
at great length. Many results were first obtained for waves of
invasion, noisy versions of the classical ``FKPP'' model by Fisher,
Kolmogorov, Petrovskii, Piskunov~\cite{van2003front}. In recent years,
however, there has been a particularly strong research focus on models
of adaptation. These models aroused widespread interest because they
can be applied to several types of data, including genomic data
derived from experimental evolution experiments and from natural
populations that undergo rampant adaptation, such as bacteria and
viruses~\cite{neher2013genetic}. We now have analytical predictions
for a number of valuable analytical or semi-analytical results for
observables such as the mean speed, probability of fixation,
distribution of fixed mutations in the asymptotic regime of large
populations~\cite{neher2013genetic}. The great value of these results
is that they show and rationalize which parameter combination chiefly
influence the dynamics, and through which functional
form. Importantly, it has been generally found that the overall
dynamics depends logarithmically on population size and mutation
rate. The weak functional dependence is in fact at the root of
universality observed in many such wave models: These predictions are
independent of the precise details of the models, including the form
of the non-linear population size regulation.

The basic challenge in analyzing noisy traveling arise from an
essential non-linearity that is required to control population
growth. Ignoring such a dynamical control of the population size leads
to long-term exponential growth or population extinction. Progress in
describing the \emph{mean} behavior of front-sensitive models has been
achieved by at least three different approaches: One can either
heuristically improve the mean-field dynamics by setting the net
growth rate equal to zero in regions where the population densities
are too small~\cite{tsimring1996rna}. Such an ad-hoc approach, based
on a growth rate cut-off, correctly reproduces the wave speed to the
leading order but does not reveal other universal next-to-leading
order corrections or the wave diffusion constant. One can also invoke
a branching-process approximation for the tip of the wave, thereby
neglecting effects of the non-linear population size control, and then
match this linearized description with a deterministic description of
the bulk of the
population~\cite{rouzine2003solitary,desai2007beneficial,rouzine2008traveling,schiffels2011emergent,good2012distribution}. Finally,
there is also the possibility to invoke a particular dynamical
constraint with the property that the dynamics exhibits a closed
linear equation for the first moment. Importantly, this method, which
has been called ``model
tuning''~\cite{hallatschek2011noisy,good2012distribution,geyrhofer2013stochastic},
reproduces the universal features of noisy traveling waves, which are
independent of the chosen population control, ultimately because of
the weakness of the population size dependence.

While understanding the mean behavior of noisy traveling waves has
been an important achievement, the actual stochastic dynamics is
characterized by pronounced fluctuations at the population level. No
two realizations of an evolution experiment, for instance, will exhibit
the same time-dependent fitness distribution because of
the chance effects involved in reproduction and mutations. Measuring
the mean behavior requires many replicates in which the entire
environment is accurately reproduced. Even if one has access to many
replicates, as is possible in highly parallelized well-mixed evolution
experiments, one can potentially learn a lot from the variability
between replicates. Thus, a predictive understanding of the
variability in evolutionary trajectories would greatly improve our
quantitative understanding of how evolution works. 

Some exact results on fluctuations at the population level
are available for a special model of FKPP waves
\cite{brunet2006phenomenological}. Still, we currently lack a systematic
approach that can be applied to a wide range of models. Here, we fill
this gap by extending the method of ``model
tuning''~\cite{hallatschek2011noisy} to the analysis of higher
correlation functions: We show that it is possible to choose a
constraint in such a way that the hierarchy of moments is closed at any
desired level. The resulting linear equations can be solved
numerically and are amenable to asymptotic analytical techniques. As an
important application of this approach, we show how the coalescence
time can be computed within traveling wave models.

Although our main results are applicable to a wide range of models, we
focus our attention on simple models of adaptation. Beyond simply
grounding our discussion, there are two reasons to focus on these
models. On the one hand, models of adaptation are simply important and
have become an indispensable tool as a null model for evolutionary
dynamics in microbial population. On the other hand, models of
adaptation manifestly exhibit a particular mathematical structure,
which we call \emph{constrained branching random walks}. As we will
argue below, this mathematical structure, to which all our formal
results apply, can be identified as the essence of a wide range of
models arising in physics, ecology and evolution.


\section{Models of adaptation as Constrained Branching Random Walks}
\label{sec:model-definition}
Darwinian adaptation spontaneously emerges from the processes of
mutation, reproduction and competition, and these features need to be
mirrored in any model of adaptation. In models, spontaneous mutations
can be represented by a stochastic jump process in a ``fitness
space''. Reproduction is naturally described by a branching process by
which individuals give birth at certain fitness-dependent
rates~\cite{allen2003introduction,haccou2005branching}. In combination,
reproduction and mutations thus generate a branching random
walk~\cite{allen2003introduction,haccou2005branching}, which by itself would lead to
diverging population sizes. To avoid this unrealistic outcome, models
of adaptation also encode a constraint on population sizes to account
for the competition for finite resources. The resulting process is a
branching random walk subject to a global constraint, which we now
frame mathematically.

The state of the population at time $t$ is described by a function
$c_t(x)$ representing the number density of individuals with fitness
$x$. In this context, fitness refers to an individual's net-growth
rate in the absence of competition for resources. The population is
assumed to evolve in discrete timesteps of size $\epsilon$, which is
eventually sent to zero in order to obtain a continuous-time Markov
process. Each timestep consists of two sub-steps. The first substep
realizes reproduction and mutations and the second substep implements
competition. 

\subsection{First substep: Reproduction and mutations}
\label{sec:repr-mutat-give}

The combined effect of reproductions and mutations can be described by
the stochastic equation
\begin{equation}
  \label{eq:BRW-genotype}
  \tilde c_{t+\epsilon}-c_t=\epsilon \LO_t c_t + \sqrt{\epsilon \ov
    c_t}\,\eta_t \;,
\end{equation}
which takes the number density $c_t$ to an intermediate value
$\tilde c_{t+\epsilon}$. The term $\epsilon \LO c_t$ represents the
\emph{expected} change in density due to reproduction and
mutations. This term is linear in the number density because the
number of offspring and mutants per timestep is proportional to the
current population density. The term $\sqrt{\epsilon \ov c_t}\,\eta_t $
represents all sources of noise arising in this setup. We will now
discuss separately the precise meaning of both terms, and
give natural alternatives for their form.

The Liouville operator $\LO_t$ depends on how the mutational process
is modeled, and various examples are discussed in the following.
A particularly simple example is provided by
$\LO_t=D\partial_x^2+x-x_0(t)$, which has been used to model asexual
evolution on a continuous fitness
landscape~\cite{tsimring1996rna,hallatschek2011noisy}. Here, the
diffusion constant $D$ quantifies the fitness variance per generation,
generated by an influx of novel mutations. To account for the
notorious observation that most mutations are deleterious, a drift
term $-v_d\partial_x$ is often included.  The linear ``reaction'' term
$x-x_0$ in $\LO_t$ simply accounts for the fact that individuals with
higher growth rate $x$ grow faster. The term $x_0(t)$ refers to the
mean fitness of the population, which separates the population with a
positive net growth rate $x>x_0$ from the less fit part of the
population with $x<x_0$.

One cannot generally assume that (biased) diffusion is a good model
for discrete mutational events because of the presence of the reaction
term favoring highly fit individuals. The diffusion approximation
requires that mutation rates are higher than the typical fitness effects
of novel mutations. This may apply to rapidly
mutating organisms, such as viruses, or close to a dynamic
mutation-selection
balance~\cite{goyal2012dynamic,neher2013coalescence}. It may also
effectively apply in island models with low migration rates, where the
fitness effect of a mutation is reduced by potentially low migration
rates. But, in well-mixed populations, the
diffusion approach breaks down when beneficial mutation rates are much
smaller than their typical effect, which has been confirmed for a
number of microbial species when they adapt to new
environments~\cite{perfeito2007adaptive,gordo2011fitness,levy2015quantitative}.  More
generally, asexual adaptation may, therefore, be cast into the form
\begin{equation}
  \label{eq:2}
  \LO_t=\mathcal{M}_t+x-x_0(t) \;,
\end{equation}
where the mutational process is described by the operator
$\mathcal{M}_t$, which conserves particle numbers, i.e., describes a
pure jump process. For instance, one may have one of the time-independent
kernels 
\begin{equation}
  \label{eq:mutational-operator}
  \{\mathcal{M}_tc_t\}(x) \sim \left\{
\begin{matrix} D \partial_x^2c_t(x)\;, & \text{Diffusion Kernel} \\    \mu\left[c_t(x-s)-c_t(x) \right]\;, & \text{Staircase Kernel} \\\int
  \mu(y) \left[ c_t(x-y)-c_t(y) \right] dy\;,  & \text{General Mutational
    Kernel}\end{matrix}
 \;\right.
\end{equation}
Here $\mu$ is a mutation rate and $s$ is a characteristic scale for
the mutational effect. The diffusion kernel is the simplest of these
kernels because it is characterized by only one compound parameter,
the diffusion constant $D=\mu s^2$, rather than two in the Staircase
Kernel or an entire function $\mu(y)$ in the general case.

The stochastic term $\sqrt{\ov\epsilon c_t}\eta_t$ in equation
\eqref{eq:BRW-genotype} accounts for all random factors that influence
the reproduction process. The function $\eta_t(x)$ represents standard
white noise, i.e. a set of delta correlated random numbers,
\begin{equation}
  \label{eq:eta-eta-corrs}
  \overline{ \eta_t(x)\eta_{t'}(y)}=\delta(x-y)\delta_{tt'}\;,
\end{equation}
where $\overline{f}$ denotes the ensemble average of a random
variable $f$, and $\delta(x)$ and $\delta_{ij}$ are the Dirac delta
function and the Kronecker delta, respectively. The amplitude $\propto
\sqrt{\epsilon\ov c_t}$ of the noise term in Eq.~
\eqref{eq:BRW-genotype} is typical for number fluctuations: Due to the
law of large number, the expected variance in population numbers from
one timestep to the next is proportional to the number $\epsilon c_t$
of expected births or deaths during one timestep. The numerical
coefficient $\ov$ is the variance in offspring number per
individual. For instance, when we assume that offspring have nearly
matching division and death rates one finds $\ov=2$.  The variance in
offspring number is typically assumed to be of order one, but could
become much larger if offspring distributions are highly-skewed, as it
is the case when few individuals produce most of the
offspring~\cite{hedgecock2011sweepstakes}. 

\subsection{Second substep: Population size constraint}
\label{sec:second-subst-comp}
Because the reproduction step generally changes population numbers,
another sub-step, following the branching process, is required to
enforce a constant population size~\footnote{Note that if $x_0(t)$
  denotes the \emph{mean fitness} of the population, the action of the
  Liouvillean $\mathcal L$ does not change the \emph{expected} number
  of individuals in the population.  However, fluctuations in the
  reproduction implemented by genetic drift will result in slight
  deviations from this expected outcome. These deviations accumulate
  over time and either lead to extinction or an ever increasing
  population size.},
\begin{equation}
  \label{eq:pop-size-fixed}
  1=\int_x \frac{1}{N}c_t(x)\;.
\end{equation}
In most models and experiments~\cite{gerrish1998fate,desai2007speed},
this step is realized by a random culling of the population:
individuals are eliminated at random from the population until the
population size constraint is restored. Mathematically, the population control step can
be cast into the form
\begin{equation}
  \label{eq:random_culling}
  c_{t+\epsilon}=\tilde c_{t+\epsilon}(1-\lambda)\;,
\end{equation}
where $\lambda$ represents the fraction of the population that has to
be removed to comply with the population size constraint. The second
sub-step completes the computational timestep, and takes the
concentration field from the intermediate state $\tilde
c_{t+\epsilon}$ to the properly constrained state
$c_{t+\epsilon}$. 

The above standard model of adaptation with fixed population size
represents a branching random walk subject to the constraint that the
total population size is fixed. Enforcing this constraint leads to the
non-linearity that makes the associated model difficult to solve.

Note that, in the above formulation, it is assumed that all noise
comes from birth-death processes.  We have ignored, for simplicity,
additional sources of noise due to, e.g. the mutational jump
processes, which are sub-dominant in large populations.

\subsection{Generalization to arbitrary linear constraints}
\label{sec:beyond-fixed-popul}
While it is necessary to constrain the population dynamics to avoid an
exponential run-away, there is no particular biological reason to
strictly fix the population size - in fact, most population sizes
fluctuate over time~\cite{frankham1995effective}. As we will see,
there are, however, mathematical reasons to consider constraints of
particular form, which greatly simplify the analysis.

As a key step towards these tuned models, we note the fixed population
size constraint Eq.~\eqref{eq:pop-size-fixed} can be viewed as one
member of a whole class of linear constraints,
\begin{equation}
  \label{eq:generalized-constraint}
  1=\int_x u_t(x) c_t(x)\equiv \langle u_t\mid c_t\rangle\;.
\end{equation}
that one could formulate with the help of a suitable weighting
function $u_t(x)$. Any such constraint will be able to limit the
population size, and thus defines, together with the Liouvillean
$\LO$, a particular model of adaptation. Our main result
will be the observation that there are an entire set of weighting
functions for which the dynamics of the model becomes simple
(cf. Sec.~\ref{sec:main-results}).

Note that one recovers the fixed population size constraint of one
chooses the weighting function $u_t(x)$ to be a constant,
$u=N^{-1}$. For any other choice, the population size will not be
fixed. At best, one obtains a steady state with a population size
fluctuating around its mean value, $\overline N_t$, which may change
depending on the time-dependence imposed on the weighting function
$u_t$. Note that culling is does not discriminate among individuals of
different fitness. It is only the \emph{amount} of culling that
depends on the distribution and type of individuals if $u_t(x)$ is
$x$--dependent.

\section{Invasion waves as constrained branching random walks}
\label{sec:general-noisy-traveling-waves}
One advantage of using the general form
Eq.~\eqref{eq:generalized-constraint} for a global constraint is that
many types of traveling waves arising in ecology and evolution can be
cast in the same mathematical form, if an appropriate Liouville
operator and weighting function are used.

\begin{figure}
  \includegraphics[width=\columnwidth]{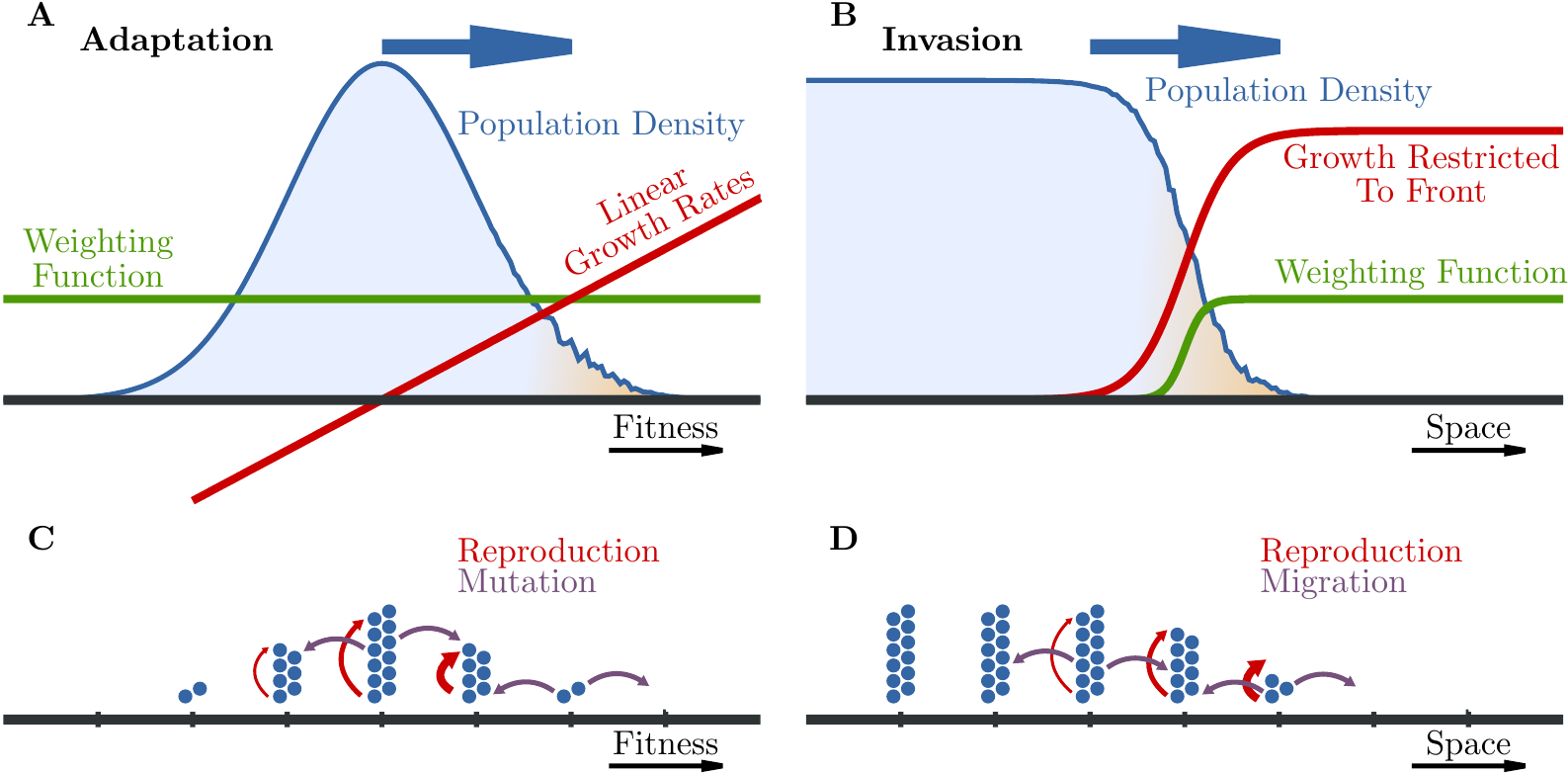}
  \caption{{\bf Models of adaptation and invasion.} Illustration of
    the essential features of two types of noisy traveling waves: {\bf
      A} Well-mixed models of asexual adaptation generate waves that
    travel across fitness landscapes towards higher fitness. {\bf B}
    Models of species invasions, range expansions or epidemics
    generate waves that propagate in real space. Both types of waves
    can be viewed as emerging from \emph{constrained branching random
      walks}. The state of the traveling wave is described by a
    (fluctuating) population density shown in blue. A branching
    process is generated by linear net-growth rates and a jump
    process. In the case of adaptive waves, individuals reproduce
    according to their current fitness and jump due to spontaneous
    mutations ({\bf C}). By contrast, the jump process in invasion
    waves is generated by random movements according to some dispersal
    kernel, and density-dependent growth rates lead to an effective
    location dependence of the growth rates ({\bf D}). A global
    constraint ensures a finite population size and depends on a
    weighting function $u_t(x)$ indicated in green. Note that for
    conventional models of adaptation with fixed population size, the
    growth rates are increasing with fitness and the weighting
    function is a constant. By contrast, for models of species
    invasion or range expansion without Allee
    effect~\cite{hallatschek2008gene}, the growth rates and weighting
    function have a sigmoidal shape, saturating in the tip of the
    wave. This ensures that individuals in the tip of the wave have
    the highest growth rates (because of least competition) and that
    their total number stays finite. }
  \label{Fig:noisy-wave-schema}
\end{figure}

We would like to give an example of this
assertion. Fig.~\ref{Fig:noisy-wave-schema}b illustrates the expansion
of a population in a real landscape, which may describe an
advantageous gene spreading through a population distributed in space,
or the invasion of virgin territory by an introduced
species. Models of such real-space waves require the
following features: i) populations reproduce and die ``freely'' in the
tip of the wave, where population densities are small, ii) individuals
move in real-space according to some jump process and iii) the
net-growth vanishes in the bulk of the wave, where resources are
sparse.

Features i) and ii) again generate a branching-random walk in the tip
of the wave, however, according to different position-dependent growth
and jump rates than in the case of evolution. Feature iii) requires a
finite population size in the tip of the wave. This non-linearity keeps
the branching-random walk away from proliferating to infinite
densities, where mean-field models apply.

To generate an adequate branching-random walk, one has to use an
appropriate Liouville operator. In one dimensions, one can choose, for
instance,
\begin{equation}
  \label{eq:Liouville-real-space}
  \LO_t=\mathcal{M}_t+s\, \Theta(x-x_0(t)) \;.
\end{equation}
Now, $x$ refers to the location in one-dimensional real space. $x_0$
refers to the position of the cross-over to the bulk of the wave. The
operator $\mathcal{M}_t$ generates a jump process. In the simplest
case, again, the jump process may be approximated by a diffusive
process, $\mathcal{M}_tc=D\partial_x^2c$. The growth term does not
need to be modeled by a strict step-function, any sigmoidal function
works in the limit of large population sizes \cite{brunet1997shift}. Importantly, the net
reproduction rates monotonously increase in $x$ and saturates at some
finite value $s$ at $x=O(1)$. Finally, to limit the population size in
the front of the wave, we need to use a non-constant weighting
function $u(x,t)$ in the global constraint, for instance,
$u_t(x)=\frac 1N \theta(x-x_0)$, which ensures that the growth region
contains precisely $N$ individuals.

The qualitative difference between the traveling waves in real and fitness
space turns out to rely on the growth rates in the nose of the wave:
While growth rates are saturating in the case of real-space waves,
it is increasing without bound in the case of waves of adaptation. This
makes models of adaptation even more sensitive to the effects of noise to the
extent that a mean-field limit (neglecting noise) does not even yield
finite velocity waves. Noise serves a crucial function in these models
as it is required to regularize the wave dynamics. These waves have
therefore been called ``front-regularized waves''~\cite{cohen2005fluctuation}.

\section{Summary of Main Formal Results}
\label{sec:main-results}
In the same way as exemplified above for models of adaptation and
invasion, one can frame many other eco-evolutionary scenarios, in
their essence, as constrained branching random walks. These models
are, ultimately, defined by an operator $\LO_t$ generating the
branching-random walk and a weighting function $u_t$ defining the
global constraint. In this paper, we show that, in fact, for any given
$\LO_t$ there are \emph{natural} ways of choosing the weighting
function. The associated models, which we call \emph{tuned} models,
have desirable properties, including closed and linear moment
equations, greatly facilitating their analysis.

We first state our main results on how to construct tuned models at
any desired level of moments. We will then provide an interpretation
of these tuned models and provide simulation results, which we compare
to fixed population size models. Detailed analytical derivations are
given in later sections.
 
To characterize fluctuations in the makeup of the population it is
convenient to consider the so-called $n$-point correlation function
$\cna$, which is the noise-average of the product
\begin{eqnarray}
  \label{eq:NPointCorrFctDef}
  C^{(n)}(x_1,\dots,x_n;t)&\equiv&\prod_{j=1}^{n}c(x_j;t)\;,
\end{eqnarray}
of $n$ number density fields, $c(x_j;t)$, evaluated at the same time
$t$ at various locations ${x_j}$. Note that, here and henceforth, we use the
notations $f_t(x)$ and $f(x;t)$ for a space and time-dependent
function interchangeably. 

From many studies over the last 15 years, we know a lot about the
first moment, $\cna[1](x;t)$, of several models of noisy
traveling waves. This provides access to the \emph{expected} shape and
velocity of the traveling wave, as well as the \emph{expected} fate of
individual mutations, or sub-populations. However, wave shape and
velocity fluctuations, as well, as the decay of genetic diversity
requires access to higher moments, $n>1$, for which there is no
systematic approach so far.

Our main result is that the dynamics of the \nth moment (and of all
lower moments) becomes analytically accessible if one
chooses in the global constraint
Eq.~\eqref{eq:generalized-constraint} the weighting function $u_t$ to
have the form 
\begin{equation}
  \label{eq:n-weighting-fct}
  \un(x;t)=\frac{2\us(x;t)}{\ov(n+1)}
\end{equation}
for any positive integer $n$, where $\us(x;t)$ satisfies
\begin{equation}
  \label{eq:selection-function-nth-moment-mr}
  -\partial_t\us=\left[\LO^\dagger +\G(t) - \us\right]\us 
\end{equation}
with the adjoint operator $\LO^\dagger $ of $\LO$. The arbitrary
function $\gamma(t)$ controls the mean total population size as a
function of time.

For the special weighting function Eq.~\eqref{eq:n-weighting-fct}, the
equation of motion for the \nth moment becomes \emph{closed} and
\emph{linear},
\begin{equation}
  \label{eq:closed-hierarchy}
  \partial_t \cna=\sum^n_{j=1} \left. \left(\LO+\gamma(t)-\frac{2n}{n+1} \us
    \right)\right|_{x_j}
  \cna+\frac{2}{n+1}\sum_{j=1}^{n}\sum_{k=
    j+1}^{n}\delta(x_j-x_k)\langle \us\mid \cna
  \rangle_{k} \;.
\end{equation}
The resulting models may be called \emph{tuned}, because for all other
choices of the weighting function, the equation of motion for
$\cna$ involves $\cna[n+1]$ resulting in an
infinite hierarchy of moments. The moment closure of the tuned models
is exact and not due to a truncation or approximation of higher
moments. The two terms $\propto \us$ are fluctuation-induced
terms. The last, positive term generates correlations at equal space
arguments, which are then dissipated by the first, strictly negative, term over longer
time scales. The positive term exists only for $n>1$ and indicates
that the effect of genetic drift on higher moments is more complicated
than a cut-off in the growth term. 

Moreover, examining the behavior of differently labeled, but otherwise
identical, subpopulations shows that tuned models can be interpreted
naturally in the framework of population genetics. First, the weighting
function of all tuned models is a fixation probability function: The
probability that descendants of individuals at location $x$ and time
$t$ will take over the population on long times is given precisely by
$\un(x;t)=2\us(x;t)/\ov(n+1)$ in the model tuned for the \nth moment. It
is remarkable, in this context, that equation
Eq.~\eqref{eq:selection-function-nth-moment-mr} for $\us(x;t)$ is
precisely the equation governing the survival probability of an
\emph{unconstrained} branching random walk~\cite{fisher2013asexual}, with $\ov=2$. 

Secondly, the higher moments provide access to the statistics of the
genetic makeup of the population. According to the principle tenet of
population genetics that, without mutations, the genetic
diversity of a population decreases with time: fixation and extinction
of subtypes needs to be maintained by an appropriate
influx of mutations. The decay of genetic diversity fundamentally
depends on higher moments -- the first moment captures fixation
probabilities but not the time to fixation.

The decay of genetic diversity in the absence of mutations can be quantified by the
cross-correlation function
\begin{eqnarray}
  \label{eq:NPointCorrFctDef-subtype}
  \cna_{i_1,\dots,i_n}(x_1,\dots,x_n;t)&\equiv&\overline{\prod_{j=1}^{n}c_{i_j}(x_j;t)}\;,
\end{eqnarray}
between subtypes $i_1,\dots,i_n$. Here, the number densities
$c_{i}(x;t)$ is the number density field of type $i$. The sum of all
sub-types makes up the total population, $c(x;t)=\sum_ic_i(x;t)$.

If all $i_j$ are different and $n>1$, it is clear that
$\cna_{i_1,\dots,i_n}$ must continuously decay in the absence of
mutations because one of the subtypes will take over in the presence
of random genetic drift. The case $n=2$ is the most prominent one:
$\cna[2]_{1,2}(x,x;t)$ is proportional to the so-called
heterozygosity at location $x$ and $t$, which is the probability that
two individuals sampled with replacement are of different
type~\cite{wakeley2009coalescent}.

Now it turns out that, in the \nth-tuned model, the cross-correlation
function can be expressed in the simple form
\begin{equation}
  \label{eq:sep-ansatz-0}
  \cna_{i_1,\cdots,i_n}=\prod_{k=1}^n f_{i_k}(x_{i_k};t)
\end{equation}
with $f_{i_k}(x_{i_k};t)$ satisfying a one-dimensional linear equation
\begin{equation}
  \label{eq:sep-ansatz-1}
  \partial_t f_{i_k}(x;t)=\left(\LO+\gamma(t)-\frac{2n}{n+1} \us
    \right) f_{i_k}(x;t)
\end{equation}
subject to the initial conditions $f_{i_k}(x;0)=c_{i_k}(x;0)$.

Notice that the decay of genetic diversity in the \nth-tuned model is
described by the spectrum of the operator appearing on the right-hand side of
Eq.~\eqref{eq:sep-ansatz-1}, which also occurs in
Eq.~\eqref{eq:closed-hierarchy} describing the fluctuations of the total
population. Consistent with our population genetic interpretation, one
can show quite generally that this operator only has decaying relaxation
modes (see Fig.~\ref{fig:spectrum-SI}).

We consider Eqs.~\eqref{eq:sep-ansatz-0} and \eqref{eq:sep-ansatz-1} to be our
results with the most immediate applicability because they provide a
feasible approach to resolving a key question in population genetics
(maintenance and decay of genetic diversity) that \emph{relies} on having access to higher moments.

\subsection{Remarks}
\label{sec:remarks}
The above formalism includes the case $n=1$ (closed first moment)
presented in Ref.~\cite{hallatschek2011noisy} (in which the tuned
weighting function $u^{(1)}$ was denoted by $u_*$).

In constrained random walk models, one can generally retrieve lower
moments from higher moments by contraction with the weighting function
$u$,
\begin{eqnarray}
  \label{eq:contracting}
   \langle u\mid C^{(n)} \rangle_{x_k}&
   =&c(x_1;t)\dots \left(\int_{x_k}u(x_k;t) c_t(x_k;t)\right)\dots c(x_{n};t) \nonumber\\
   &=&c(x_1;t)\dots c(x_{k-1};t) \,1 \, c(x_{k+1};t)\dots c(x_{n-1};t)\nonumber \\
   &\equiv& C^{(n-1)}(\backslash x_{k};t) \;,
\end{eqnarray}
where we invoked the global constraint
Eq.~\eqref{eq:generalized-constraint} in going from the second to the
third line. To simplify the notation, we have here introduced the short hand
$C^{(n-1)}(\backslash x_m;t)$ to denote
$C^{(n-1)}(x_1,\dots,x_{m-1},x_{m+1},\dots,x_n;t)$, i.e., that the
$m$th space variable should be omitted in the arguments. 

The contraction rule Eq.~\eqref{eq:contracting} also shows that contraction
in the last term in the dynamical equation for $C^{(n)}$ simply
generates the next lower moment $C^{(n-1)}$. Moreover, if one obtains
the \nth moment by solving Eq.~\eqref{eq:closed-hierarchy}, one
generally has access to all lower-rank moments as well (but not to the
larger moments). Thus, the model tuned to be linear at the \nth moment
gives access to all moments up to and including the \nth.

It is remarkable that the governing equations for $\us$ and
$\cna$ are independent of the offspring number variation
$\ov$. The only effect of $\ov$ is to reduce the fixation probability $\un\propto \ov^{-1}$ and
scale up the population sizes $\cna\propto
\ov^{n}$. This means in particular that noise-induced terms $\propto\us$ in the
moment equation are not scaled by $\ov$, and thus do not become small
in the small noise limit. The lack of a potentially small parameter
has important consequences for the use of perturbation theory in this
context.

\section{Numerical Results}
\label{sec:numerical-examples}
We now illustrate our main results by explicit numerical solutions and
stochastic simulations for models tuned to be closed at the first and
second moment. The two biological phenomena we consider, adaptation and invasion,
both give rise to compact traveling waves in real space and fitness
space, respectively. They fundamentally differ
in their location-dependent growth rates: While in adaptation models,
growth rates are linear increasing towards the tip of the waves, they
saturate in invasion models. As a consequence, invasion waves have a
well-defined infinite population size limit in contrast to adaptation
waves.

\subsection{Models of adaptation}
\label{sec:models-of-adaptation}
The detailed behavior of simple models of asexual adaptation models
varies with the assumptions made about how the mutation process
influences the growth rates that define the branching process. In our
numerical work, we focus on the diffusion kernel in
Eq.~\eqref{eq:mutational-operator}, which assumes that a growth rate
variance $D$ is acquired per generation due to mutations. The
advantage of the diffusion kernel is that it only contains one
parameter, the diffusion constant $D$, and matters when mutation rates
are large compared to the (effective) rate of
selection~\cite{tsimring1996rna,goyal2012dynamic}.

\paragraph{Numerical Approach.} To solve our framework of tuned
models, we used a multi-dimensional Newton-Raphson method to determine
traveling wave solutions corresponding to tuned models of degree
$n$. To this end, we first determined a traveling steady state
solution $\us(x,t)=\us(x-vt)$ of
Eq.~\eqref{eq:selection-function-nth-moment-mr} defining the set of
tuned weighting function $\un(x,t)=2\us(x,t)/b(n+1)$
(Eq.~\eqref{eq:n-weighting-fct}). This weighting function enforces a
traveling steady-state for the \nth moment $\cna(x_1,...,x_n;t)$ of the
number densities, which satisfies the linear time-independent equation
Eq.~\eqref{eq:closed-hierarchy}. As a result, we obtain for a given $n$, a weighting function
$\us(x)$ and the \nth moment. Details of the numerical scheme and the
explicit equations solved for $n=1$ and $n=2$ are described in Appendices \ref{appendix:explicit} and \ref{appendix:numerics}.

\begin{figure}[tb]
\begin{center}
\includegraphics[width=\textwidth]{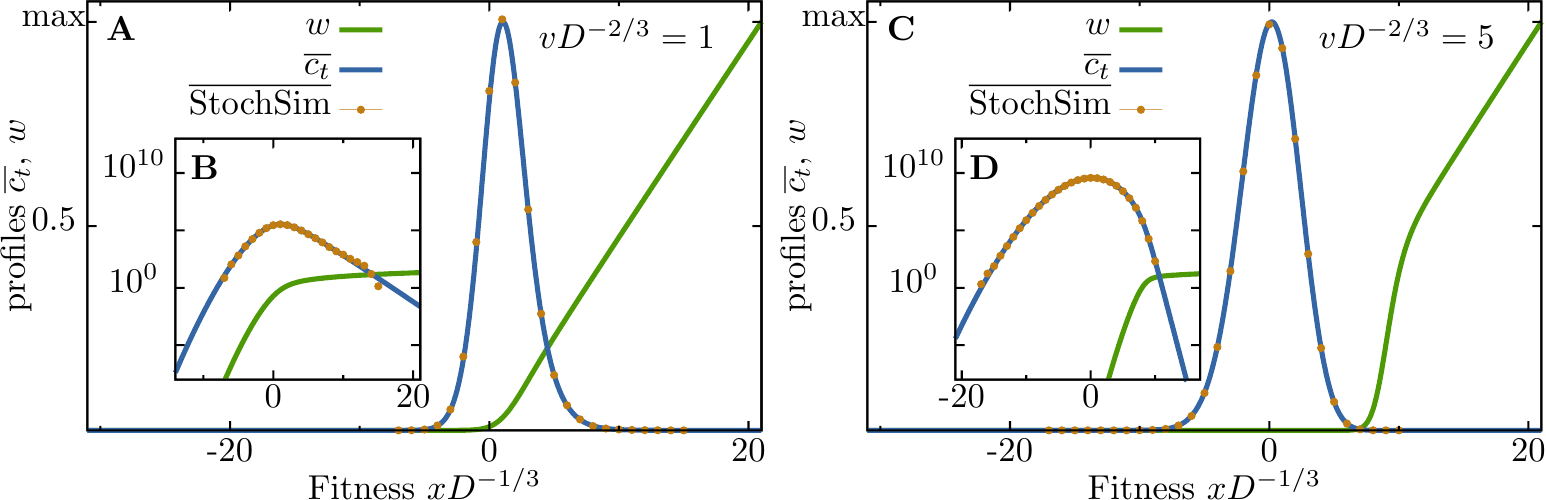}
\caption{\label{fig:numericalsolution} \textbf{Numerical solution of
    the model of adaptation tuned to be closed at the first moment
    ($n=1$), and comparison with stochastic simulations.}  Averages
  from stochastic simulations (brown) coincide with numerical solutions of the mean stationary population
  density $\overline{c_t}$ (blue line) for small speeds ({\bf A, B}) and
  large speeds ({\bf C, D}), respectively.  The weighting function
  $w(x,t)$ (green line) exhibits a sharp increase towards the nose of
  the wave, before crossing over to its asymptotic linear solution,
  $w(x,t)\approx x$, for large fitness. This is consistent with
  the interpretation of $\un(x,t)=w(x,t)/(n+1)$ as fixation probability, tested in
  Fig.~\ref{fig:measurefixation}. Models tuned to be closed at higher
  moments ($n>1$) display the same features qualitatively and
  quantitatively, see Fig.~\ref{fig:highermoments}.  \ }
\end{center}
\end{figure}
\begin{figure*}[tb]
\begin{center}
\includegraphics[width=\textwidth]{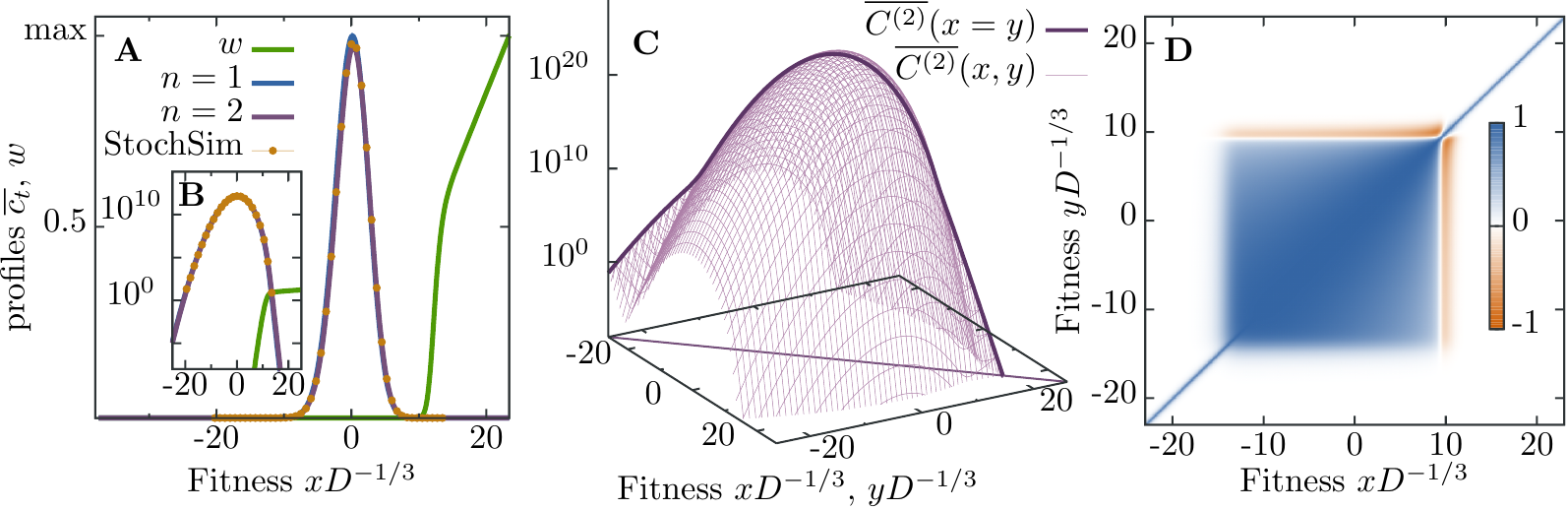}
\caption{\label{fig:highermoments} \textbf{Numerical solution of the
    model of adaptation tuned to be closed at the second moment
    ($n=2$).}  {\bf A, B} The mean population densities of the models
  with $n=2$ (purple) and $n=1$ (blue) for the same wave speed (
  $vD^{-1/3}=6$) almost coincide. The profile for $n=2$ (and generally
  for larger $n$) is shifted to slightly larger fitness values, but
  the population size $\overline{N}$ is almost unaffected
  ($\overline{N}^{(n=1)}/\overline{N}^{(n=2)}-1 \lesssim 0.02$), with
  even less deviations for larger population sizes (or adaptation
  speeds).  Profiles from stochastic simulations subject to the
  constraint defined by $u^{(2)}$ are shown as brown dots. The
  standard error is smaller than the dot size. {\bf C} The numerical
  solution of the second moment $\cna[2](x,y)$ ($n=2$) is
  shown in a semi-logarithmic plot. The ridge $\cna[2](x=y)$
  is nearly parabolic except for its exponential tails.  {\bf D} From the
  second moment, we calculate the Pearson-product-moment-correlation
  $\rho(x,y)$ (see main text), which allows us to distinguish
  correlated (+1), uncorrelated (0) and anticorrelated (-1)
  variation. The nose of the wave clearly is anticorrelated with the
  bulk of the wave in our tuned models.  }
\end{center}
\end{figure*}

\paragraph{Model tuned to the first moment ($n=1$).} To set the
stage and to reproduce earlier results from
Ref.~\cite{hallatschek2011noisy}, we first present numerical results
for the model closed at the first
moment. Fig.~\ref{fig:numericalsolution} shows, for two wave speeds,
the weighting function and the mean population density in a stationary comoving frame. While the
population distribution is, except for an exponential decay in the
wave-tip, close to a Gaussian for large speeds, it is markedly skewed
for lower wave speeds. Note that the exact numerical results are in
near perfect agreement with  stochastic simulations, confirming our
approach. 

\paragraph{Model tuned to the second moment ($n=2$).}
Fig.~\ref{fig:numericalsolution} characterizes the behavior of $n=2$
models in comparison with $n=1$ tuned models. Since $n=2$ models
provide access to the second moment, one can of course also obtain the first, by
contraction with the weighting function $\un(x,t)$
(cf. Eq.~\eqref{eq:closed-hierarchy}). Fig. ~\ref{fig:numericalsolution}B shows the mean
population densities for both $n=1$ and $n=2$ models for the same
velocity and weighting function $\us$ which is identical to their respective
weighting functions $u^{(1)}$ and $u^{(2)}$ up to an $n$--dependent
factor, cf. Eq.~\eqref{eq:n-weighting-fct}. The mean population
densities can hardly be distinguished in the figures - generally, the
agreement of the two different models increases with increasing wave
velocity or, equivalently, population size. Moreover, the stochastic
simulations track the predicted mean in near perfect agreement,
providing numerical support for our analysis of the tuned closure at
higher moments.

Fig.~\ref{fig:numericalsolution}B shows, in 3d plot, the second moment
$\cna[2](x,y)$ in a semi-logarithmic plot. While this plot is mostly
Gaussian, it reveals a distinctly exponential decay in the front and
back of the wave. This indicates the importance of fluctuations in
these low-density regions.

Higher order correlation functions can also be used to directly
investigate fluctuation properties of the noisy adaptation wave.  The
Pearson-product-moment-correlation, defined as $\rho(x,y) =
\mathrm{Cov}[c_t(x),c_t(y)]/\sqrt{\mathrm{Var}[c_t(x)]\mathrm{Var}[c_t(y)]}$
shows a clear anticorrelation signal in the nose of the wave: Usually,
a (stochastic) rise in population number in the nose of the wave leads
to an overall decrease in population size.  Such a very fit population
has a much larger weight in the tuned constraint,
Eq.~\eqref{eq:generalized-constraint}, forcing the bulk population to be culled.
Thus, bulk population size and nose population size are anticorrelated
\cite{geyrhofer2015oscillations}.

\paragraph{Fixation probabilities.}

Note from Figs.~\ref{fig:numericalsolution}
and~\ref{fig:highermoments} that the weighting functions
$\un(x,t)=2\us(x,t)/b(n+1)$ strongly increase towards the tip of the
wave where it crosses over to a linear increase. The functional form is
consistent with the interpretation of a fixation probability: The
success probability of an individual should be much larger in the tip
of the wave rather than the bulk because it has to compete with less
and less equally or more fit individuals. The fixation probability
approaches a linear branch beyond some cross-over fitness because
individuals there are so exceptionally fit that they merely have to
survive random death to fix. The competition with conspecific is
minimal - there is only competition with their own offspring.

\begin{figure}[tb]
\begin{center}
\includegraphics[width=.5\textwidth]{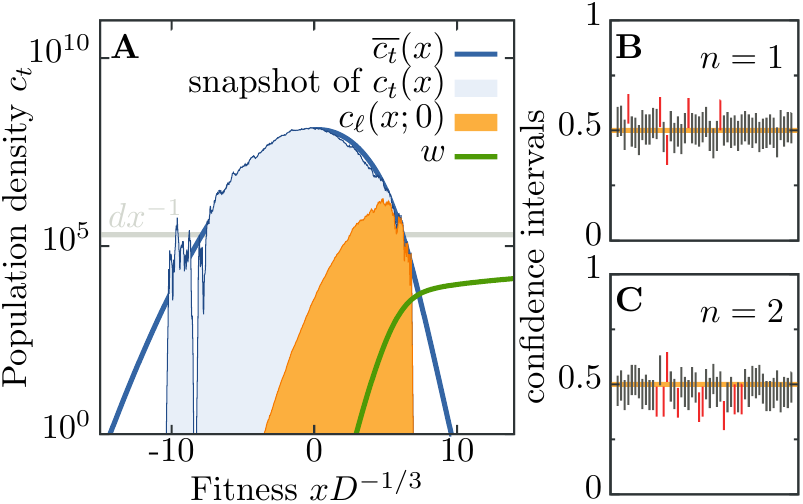}
\caption{\label{fig:measurefixation} \textbf{Fixation in traveling
    wave models of adaptation sensitively depend on individuals in the
    nose of the wave.}  One of our key results is that tuned models
  generally have the convenient feature that fixation probabilities
  can be computed exactly -- they are given by the tuned weighting
  function $\un(x,t)=2w(x,t)/\ov(n+1)$. Here, we provide a test of this
  prediction. First, we have generated $50$ independent start
  configurations, and labeled the wave tip such that the labeled
  population is predicted to have a $50\%$ fixation probability,
  $\bigl\langle u^{(1)}\vert c_\ell\bigr\rangle = 0.5$. Panel {\bf A} shows
  one start configuration and the labeled subpopulation in
  orange. (The shown density profile (light blue area) is momentarily
  smaller than the mean density (blue line) due to random population
  size fluctuations, cf. Fig.~\ref{fig:popsizespeed}).  For each of
  the labeled start configurations, we ran $200$ simulations up until
  the (fixation) time where either the labelled subpopulation is
  extinct or has taken over.  From counting the number of extinction
  and fixation events, Bayesian inference allows to compute the
  posterior distribution assuming a flat prior
  \cite{geyrhofer2014quantifying}. {\bf B} For the model with $n=1$, we
  find that $45$ of the $95\%$-confidence intervals of this
  Beta-distributed posterior include the expected fixation probability
  of $1/2$ (black intervals), while $5$ intervals scatter considerably due
  to lattice effects in simulations (red intervals). {\bf C} When closing the
  moment hierarchy at $n=2$ results exhibit stronger scattering. 
  However, the measured confidence intervals for $n=2$ reproduce the
  expected outcome to reasonable extent.  }
\end{center}
\end{figure}

To test our prediction that $\un(x)$ can be interpreted, exactly, as
an fixation probability, we have carried out the following test: We ran
simulations for $n=1$ and $n=2$ tuned models, in which we labeled the
tip of the wave such that the predicted fixation probability is
exactly $50\%$, cf. Fig.~\ref{fig:measurefixation}. The measured fixation probability is shown in the
insets of Fig.~\ref{fig:measurefixation}. Within the statistical error, the agreement is very good.

We would like to point out that the fact that $\us(x,t)$ is strongly
increasing towards the wave tip means that highly fit individuals have
a large impact on what fraction of the population is culled per time
step. For instance, if a single individual arises far out in the tail of
the wave, it may have a large fixation probability. To keep the total
fixation probability at one, this means that a significant
fraction of the individuals need to be cleared from the
population. Importantly, however, the culling itself is independent of
the identity of individuals, i.e., a poorly fit individual is equally
likely to die as a highly fit individual. This distinguishes our model
from models that regularize the population size by removing cells
preferentially at the wave tip. Such a procedure strongly modifies the
wave dynamics and, in particular, reduces the amount of
fluctuations.

\begin{figure}[tb]
\begin{center}
\includegraphics[width=0.5\textwidth]{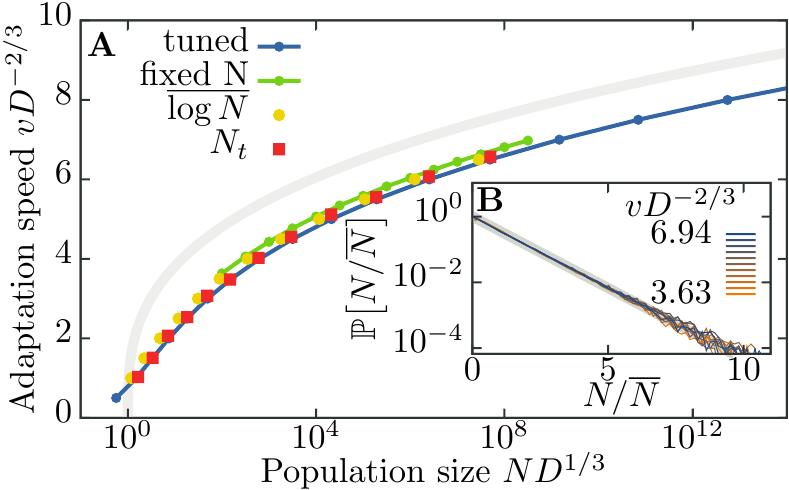}
\caption{\label{fig:popsizespeed} \textbf{Fluctuations in adaption
    waves suggest the adaptation speed is controlled by the \emph{mean
      logarithmic population size}.}  {\bf A} Stochastic simulations
  reveal that, for a given speed of adaptation, tuned models have a
  larger mean population size $\overline{N}$ (blue line) than models
  of fixed population size (green line). The small discrepancy is much
  lower between these two models than to the best available cutoff
  theory (gray line \cite{cohen2005front}).  Most of the remaining
  discrepancy is due to population size fluctuations (very nearly
  exponentially distributed, see inset {\bf B}): When we measure fluctuations
  in a stochastic realization of the tuned model and impose exactly
  the same population size time series in repeated stochastic
  simulations, we obtain adaptation speed vs. mean population size
  relation that coincides with the one obtained for the tuned models
  (red circles). Importantly, fixed population size and tuned models
  have almost the same mean logarithmic population size,
  $\overline{\log N}$, for the same wave speeds. It can be shown that
  this is due to a time scale separation between slowly decaying
  population size fluctuations (coalescence time scale) and the fast
  relaxation of the wave speed (determined by a mixing time in the tip
  of the wave). Thus, the key dynamical quantity in traveling wave
  models with fluctuating population sizes is $\overline{\log N_t}$
  instead of $\overline{N_t}$.}
\end{center}
\end{figure}

\paragraph{Velocity-Population-Size Relationships} Fig.~\ref{fig:popsizespeed} shows the relation between velocity and mean population size in
various tuned models. As comparison, we also show the corresponding
relationships between mean velocity and population size for fixed
population size models. All these models correspond to different
statistical ensembles, yet for large population sizes the curves
approach each other very well. 

The differences between models at finite population sizes is explained
mainly by population size fluctuations: If we force the fixed
population size model to fluctuate precisely as a given realization of
a tuned model, we obtain very accurate agreement. Conversely, if we
plot velocity vs. $\overline{\log N}$ shows excellent agreement
between all models. This agreement is due to a timescale separation,
discussed below.

\begin{figure}[tb]
\begin{center}
\includegraphics[width=0.5\textwidth]{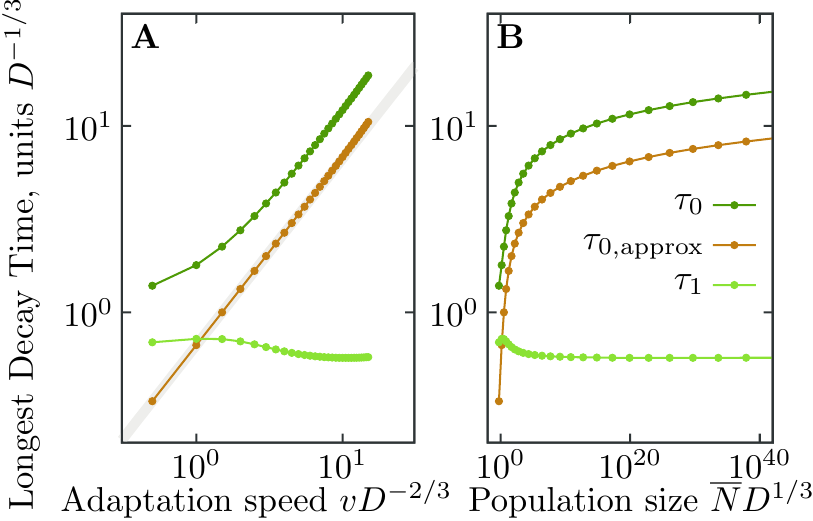}
\caption{\label{fig:decaytimes} \textbf{Time scales for the decay of
    genetic diversity in models of adaptation.} In models with limited
  population size, the ultimate fate of a subpopulation of neutral
  mutants is to either go extinct or to fixation. Hence, the diversity
  of the population will gradually decay unless new mutations come in,
  and quantifying this decay is an important population genetic
  challenge. This figure quantifies the leading decay times of the
  function $\cna[2]_{1,2}$ correlating the densities of
  labeled and unlabeled individuals. The decay times were found from a
  spectral analysis of the equation of motion of
  $\cna_{1,2}$, which is linear in the model tuned to be
  closed at the level of two-point correlation functions ($n=2$). {\bf A}
  Our numerical results suggest that, for the considered model with
  diffusion kernel, the slowest decay time approaches
  $\tau^{(n=2)}_0\propto v\sim \bigl(\ln N\bigr)^{1/3}$ for large speeds (gray
  line indicates $\bigl(\ln N\bigr)^{1/3}$). By contrast, the second slowest decay
  time $\tau^{(n=2)}_1$ approaches a constant for large
  populations. This indicates an important time scale separation, as
  we argue in the main text. Also note that the numerical results
  approach, for large $N$, the approximation
  $\tau_{0,\,\mathrm{approx}}$ in Eq.~\eqref{eq:heterozygosity-var-eom},
  which is based on the time scale separation of the two lowest
  eigenvalues emerging for large population sizes.  Panel {\bf B} shows
  the same quantities as Panel {\bf A} as a function of the population size
  $\overline N$.  }
\end{center}
\end{figure}

\paragraph{Decay of Genetic Diversity} Next, we have solved numerically
for the mode spectrum that governs the decay of heterozygosity. Fig.~\ref{fig:decaytimes} shows the behavior of the lowest two eigenvalues as
a function of the velocity of the wave and the population size, respectively. It can be clearly seen that a time
scale separation arises: The frequency of the first mode decays slowly
to $0$, following $\tau_c\sim v$ to a good approximation. The
frequency of the next higher mode, on the other hand, approaches a
constant value of order $1/D$.

This means that coalescence takes much longer than the time until a
subpopulation has forgotten its the initial condition of its spatial
distribution. This time-scale separation not only helps in
analytically finding the coalescence time in Sec.~\ref{sec:time-scale-separ}. But it also
underlies the ensuing Bolthausen-Sznitman coalescence in many models of
adaptation and invasion waves of the Fisher-Kolmogorov
type~\cite{desai2013genetic,neher2013genealogies}.

\begin{figure}[tb]
\begin{center}
\includegraphics[width=\textwidth]{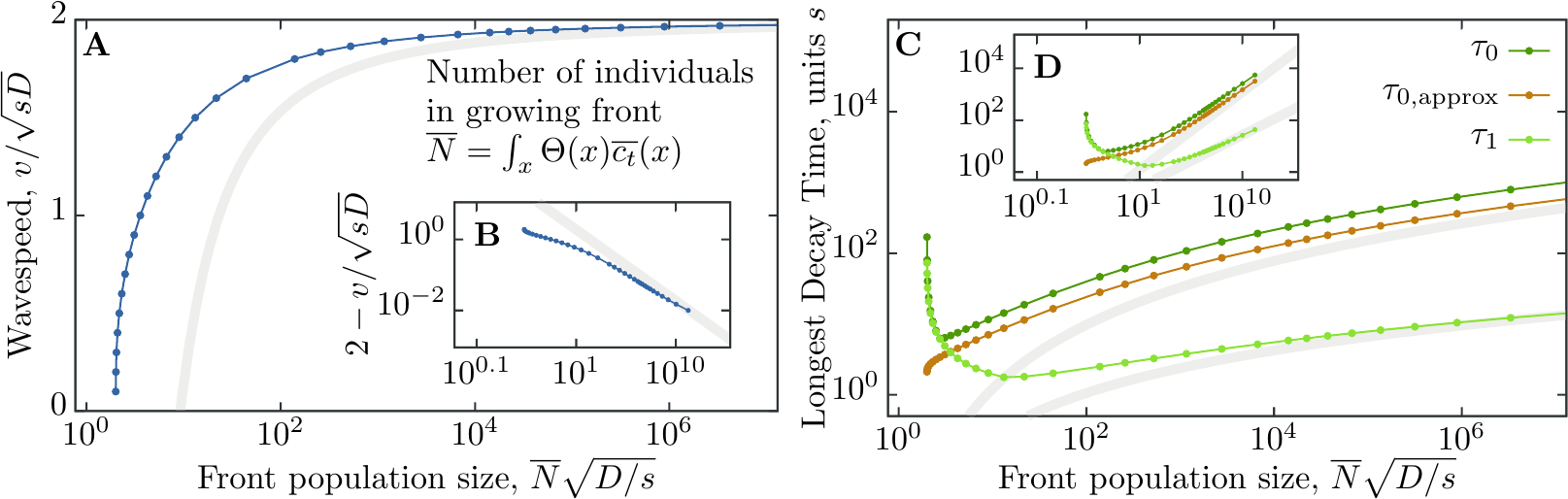}
\caption{\label{fig:fisherwaves} \textbf{Invasion waves.} Here, we
  summarize our numerical results for models of invasion tuned to be
  closed at the second moment, $n=2$. {\bf A} Wave speed as a function
  of population size. The inset {\bf B} compares our results with the
  leading order cut-off correction $v \sqrt{Dr}-2\sim \pi^2/(\ln
  \overline{N})^2$. {\bf C} shows the two longest decay times,
  $\tau_0>\tau_1$, of the second moment. The increasing gap between
  both decay times manifests a time scale separation.  The double
  logarithmic plot in inset {\bf D} shows that the longest
  decay time follows $\tau_0\sim\ln N^3$ (upper gray lines)
  asymptotically and, hence, behaves as the predicted coalescence time
  in FKPP waves~\cite{brunet2006phenomenological}. By contrast, the
  second decay time follows $\tau_1\sim\ln N^2$ (lower gray line).}
\end{center}
\end{figure}

\paragraph{Invasion waves}
\label{sec:FKPP waves}
After changing the Liouville operator the one in
Eq.~\eqref{eq:Liouville-real-space}, and following the same numerical
pipeline as described above, we obtain analogous results for invasion
waves, see Fig.~\eqref{fig:fisherwaves}. Note that the spatial
co-ordinate now corresponds to real space rather than a fitness
landscape. Our data reproduce the universal velocity-population size
relationship and the coalescence time scaling that have been
established for FKPP waves over the last 20
years~\cite{van2003front,brunet2006phenomenological}.  An explicit
form of the equations we solved numerically can be found in Appendix
\ref{appendix:explicit}.

\section{The Stochastic Dynamics of Branching Random Walks}
\label{sec:derivations}

In the following we proceed with the derivation of our main results quoted
above. To this end, we will first recapitulate the derivation of the
stochastic differential equation governing the dynamics of branching
random walks, as was done in Ref.~\cite{hallatschek2011noisy}. We will
then discuss the consequences of this stochastic dynamics for
$n$-point correlation functions, which will allow us to identify the
natural choice of the weighting function $\un$. Subsequently, we will
discuss how to modify our basic model to account for different
subtypes within the population.

\subsection{Stochastic dynamics of Constrained Branching Random Walks}
\label{sec:stochastic-dynamics}
We will now determine the stochastic dynamics obtained in the limit
$\epsilon\to0$, which is in general non-linear and hence not
solvable. We then use the ensuing stochastic differential equation to
identify special models with closed moment hierarchies. We will find
that these tuned models are not only solvable, but also allow for a
natural interpretation of the constraint in terms of fixation
probabilities.

The stochastic dynamics of a constrained branching random walk (CBRW)
was derived in Ref.~\cite{hallatschek2011noisy} and may be summarized
as follows. The state of the system is described by the number density
$c_t(x)$ of random walkers at position $x$ and time $t$. At any time,
the distribution of random walkers has to satisfy a global constraint
defined by Eq.~\eqref{eq:generalized-constraint}.  

The combination of Eqs.~\eqref{eq:BRW-genotype} and
\eqref{eq:generalized-constraint} can be written as a fraction,
\begin{eqnarray}
\label{eq:fraction} c_{t+\epsilon} &=& \frac{c_t + \epsilon\LO c_t +
  \sqrt{\epsilon\ov c_t}\eta_t}{\bigl\langle u_{t+\epsilon} \vert c_t
  + \epsilon\LO c_t + \sqrt{\epsilon\ov c_t}\eta_t \bigr\rangle} \;,
\end{eqnarray}
in the continuous-time limit (small enough $\epsilon$ is required to
ensure that the denominator of the fraction is never far from
$1$). Note that the expression in Eq.~\eqref{eq:fraction} evidently
satisfies the constraint, $\langle
u_{t+\epsilon}|c_{t+\epsilon}\rangle=1.$

Moreover, in the continuous-time limit, we only need to retain terms
up to order $O(\epsilon)$~\cite{van1992stochastic}. Thus, expanding
Eq.~\eqref{eq:fraction}, we obtain
\begin{eqnarray}
\nonumber c_{t+\epsilon} &=& c_t + \sqrt{\epsilon}\left[\sqrt{\ov c_t}\eta_t  - c_t\bigl\langle u_t\vert \sqrt{\ov c_t}\eta_t\bigr\rangle \right] \\
&& ~~+\epsilon\left[\LO c_t - \sqrt{\ov c_t}\eta_t\bigl\langle u_t\vert \sqrt{\ov c_t}\eta_t\bigr\rangle - c_t\bigl\langle\partial_t u_t\vert c_t \bigr\rangle - c_t \bigl\langle u_t\vert \label{eq:c-fractionexpansion}\LO c_t \bigr\rangle + c_t \bigl\langle u_t\vert \sqrt{\ov c_t}\eta_t \bigr\rangle^2    \right]\;.
\end{eqnarray}
In Eq.~\eqref{eq:c-fractionexpansion}, we required $u_t$ to change only
deterministically, $u_{t+\epsilon} = u_t + \epsilon\partial_t u_t
+o(\epsilon)$, i.e. it has no stochastic component.

Finally, we replace products of order $\eta_t\eta_{t'}$ in the
deterministic term of expansion Eq.~\eqref{eq:c-fractionexpansion}
with their averages, Eq.~\eqref{eq:eta-eta-corrs}~\footnote{We note
  that Eq.~\eqref{eq:expanding_corr_fct_in_eps} is a manifestation of
  Ito's rule for non-linear variable substitutions in stochastic
  differential equations.}, to arrive at the following stochastic
differential equation: The temporal change $\Delta c_t\equiv
c_{t+\epsilon}-c_t$ of the
concentration field $c_t$ from time $t$ to $t+\epsilon$ can be
written as
\begin{equation}
  \label{eq:delta-c}
    \Delta c_{t}(x) \equiv c_{t+\epsilon}-c_t=\epsilon \Deltad c_{t}(x)+ \sqrt\epsilon
\Deltas c_{t}(x)\;, 
\end{equation}
which consists of a deterministic change $\epsilon \Deltad c_{t}$ of
order $O(\epsilon)$ and a stochastic change $\sqrt{\epsilon}\Deltas
c_{t}$ of order $O(\epsilon^{1/2})$. These are given by
\begin{eqnarray}
  \label{eq:stoch-det-parts}
  \Deltad c_{t}(x)&=&\left(\LO_t- \ov u_t\right)
    c_t-\bigl\langle \bigl(\partial_t+\LO_t^\dagger- \ov u_t\bigr)u_t \mid c_t \bigr\rangle
    c_t  \nonumber\;,\\
  \Deltas c_{t}(x)&=&\eta_t\sqrt{\ov c_t}-\bigl\langle u_t \mid \eta_t
    \sqrt{\ov c_t} \bigr\rangle c_t\;.
\end{eqnarray}
Thus, we have arrived at the continuous-time stochastic process for a
constrained branching random walk~\cite{hallatschek2011noisy}, which
summarizes the combined effect of the original two-step algorithm, (i)
``branching random walk'' and ``enforce constraint''. The related
concept of forcing the solution of a SDE onto a manifold has been
analyzed in \cite{katzenberger1991solutions}.



\subsection{Moment equations}
\label{sec:moments}
This section introduces the hierarchy of moment equations of CBRWs
that characterize the mean and the fluctuations of the concentration
field of the random walkers.

Consider the dynamics of the products of $C^{(n)}_t$. Our goal is to
determine how the function $C^{(n)}_t$ changes as time marches
forward. To this end, we express the time increments of the $n$--point
products $C^{(n)}_t$ in terms of the changes of the single fields
$c_t$,
\begin{equation}
  \label{eq:n-point-time-increments}
  C^{(n)}_{t+\epsilon}=
 \prod_{j=1}^{n}c_{t+\epsilon}(x_j) =\prod_{j=1}^{n} \left[c_{t}(x_j)+\epsilon\Delta^{(d)} c_{t}(x_j) + \sqrt{\epsilon}\Delta^{(s)}c_t(x_j) \right]\;,
\end{equation}
using the deterministic and stochastic time increments $\Delta^{(d/s)}
c$ computed in \eqref{eq:stoch-det-parts}.  Next, we expand the product
  up to order $O(\epsilon)$,
\begin{eqnarray}
  \label{eq:expanding_corr_fct_in_eps}
  \Delta C^{(n)}_{t}&=&
  \sqrt{\epsilon}\sum_{j=1}^{n}C_t^{(n-1)}(\backslash x_j)\Deltas c_{t}(x_j)\\
  &&+\epsilon \left[\sum_{j=1}^{n} C_t^{(n-1)}(\backslash x_j)\Deltad
    c_{t}(x_j) + \sum_{j=1}^{n-1}\sum_{k= j+1}^{n} 
    C_t^{(n-2)}(\backslash x_j,\backslash x_k)\Deltas c_{t}(x_j) \Deltas c_{t}(x_k)\right]\;.\nonumber
\end{eqnarray}
Note that the last term arises only for $n\geq 2$. Inserting the time
increments Eq.~\eqref{eq:stoch-det-parts} of the single fields into
Eq.~\eqref{eq:expanding_corr_fct_in_eps} yields 
\begin{eqnarray}
  \label{eq:moments-1}
  \Delta C^{(n)}_t &=& \sqrt{\epsilon} \left[
    \sum^n_{j=1} \sqrt{\ov c_t(x_j)} \eta_t(x_j)C_t^{(n-1)}(\backslash x_j)
  -  n \bigl\langle u_t \mid \eta_t \sqrt{\ov c_t} \bigr\rangle  C_t^{(n)}
  \right]\\
  &&+\epsilon \left[ \sum^n_{j=1} \left.\left(\LO - \ov u_t\right)
  \right|_{x_j} C^{(n)}_t - n \bigl\langle \left(\partial_t+\LO^\dagger-\ov u_t\right)u_t \mid c_t \bigr\rangle C_t^{(n)} \right] \nonumber \\
&&+ \epsilon  \sum_{j=1}^{n-1}\sum_{k=j+1}^{n} \ov \sqrt{c_t(x_j)c_t(x_k)}\eta_t(x_j)\eta_t(x_k)C^{(n-2)}_t(\backslash x_j,\backslash x_k) 
\nonumber\\
&&-\epsilon (n-1) \sum^n_{j=1} 
  \sqrt{\ov c_t(x_j)}\eta_t(x_j)\bigl\langle u_t \mid \eta_t
    \sqrt{\ov c_t}\bigr\rangle C^{(n-1)}_t(\backslash x_j) \nonumber \\
&&+\epsilon \frac{n (n-1)}2 \bigl\langle u_t \mid \eta_t \sqrt{\ov c_t} \bigr\rangle^2 C^{(n)}_t \;.\nonumber
\end{eqnarray}
Within the deterministic $O(\epsilon)$ terms, we again replace
products of noises in terms of their averages, as given by
Eq.~\eqref{eq:eta-eta-corrs}\footnote{Note that we can always write $
  \eta_t(x)\eta_{t'}(y)=\delta_{tt'}\delta(x-y)$ plus a stochastic component. If such
  terms quadratic in the noise arise in the deterministic $O(\epsilon)$ part
  of any stochastic equation, one may simply ignore their stochastic
  component, as it would lead to (in the limit $\epsilon\to 0$) negligible
  contributions.~\cite{van1992stochastic}}. We then obtain 
\begin{eqnarray}
  \nonumber\Delta C^{(n)}_t &=& \sqrt{\epsilon}\left[\sum^n_{j=1} C_t^{(n-1)}(\backslash x_j)\sqrt{\ov c_t(x_j)} \eta_t(x_j)- n  C_t^{(n)}\langle u \mid \eta_t \sqrt{\ov c_t} \rangle \right]\\
  \nonumber &&+\epsilon\left[\sum^n_{j=1} \left. \left(\LO-\ov n u_t \right)\right|_{x_j} C^{(n)}_t +\ov \sum_{j=1}^{n}\sum_{k=j+1}^{n}\delta(x_j-x_k)C^{(n-1)}_t(\backslash x_k) \right] \\
  \label{eq:stoch-Eqn-for-Nth-moment} &&-\epsilon n\left\langle \left(\partial_t+\LO^\dagger-b(n+1)/2\,u_t\right)u_t\mid c_t\right\rangle C^{(n)}_t\;.
\end{eqnarray}
Upon averaging and sending $\epsilon$ to zero, we obtain an equation
of moment for the $n$th moment,
\begin{equation}
  \label{eq:Eqn-for-Nth-moment}
  \partial_t \cna_t=\sum^n_{j=1} \left. \left(\LO-\ov n u_t \right)\right|_{x_j} \cna_t +\ov\sum_{j=1}^{n}\sum_{k=j+1}^{n}\delta(x_j-x_k)\cna[n-1]_t( \backslash x_k) -n\bigl\langle \left(\partial_t+\LO^\dagger-b(n+1)/2\,u_t\right)u_t\mid \cna[n+1]_t\bigr\rangle_{x_{n+1}}\;.
\end{equation}
From \eqref{eq:Eqn-for-Nth-moment} we observe that the coupling to higher moments $\cna[n+1]_t$ is mediated via the dynamics of the weighting function $u_t$ in the last term.

\subsection{Exact closure}
\label{sec:Closure}
The key result of Ref.~\cite{hallatschek2011noisy} was that a
particular choice of the weighting function $u_t$ exists, for which the first
moment equation is closed. We now show that exact closures can be found for
higher moments as well. 

Suppose, the solution $u^{(n)}_t(x)$ of
\begin{equation}
  \label{eq:selection-function-nth-moment}
  -\partial_tu_t^{(n)}=\left[\mathcal L^\dagger - \frac{b(n+1)}{2} u_t^{(n)}\right]u_t^{(n)}
\end{equation}
exists, and we choose $u_t^{(n)}$ as the weighting function. For this
particular model, the dependence on the $(n+1)$th moment in Eq.~\eqref{eq:Eqn-for-Nth-moment}
disappears identically:
\begin{equation}
  \label{eq:topmoment-only-dependent-on-lower-ones}
  \partial_t \cna_t=\sum^n_{j=1} \left. \left(\LO-\ov n \un_t
    \right)\right|_{x_j}
  \cna_t +\ov\sum_{j=1}^{n}\sum_{k= j+1}^{n}\delta(x_j-x_k)\cna[n-1]_t( \backslash x_k) \;.
\end{equation}
Thus, the hierarchy of the first $n$ moments is closed. In
fact, this closed set of $n$ differential equations can be summarized
by a single integro-differential equation,
Eq.~\eqref{eq:closed-hierarchy}, because contracting $\cna_t$ with $\un$ reduces the order of the moments by virtue
of the constraint, cf. Eq.~\eqref{eq:contracting}.

The final form of our results
Eq.s~\eqref{eq:selection-function-nth-moment-mr}, \eqref{eq:closed-hierarchy} are
obtained upon substituting 
\begin{equation}
  \label{eq:weighting-fct-2}
  u_t^{(n)}(x) = \frac{2\us(x;t)}{b(n+1)}\;,
\end{equation}
which is the initially quoted Eq.~\eqref{eq:n-weighting-fct}.

In summary, starting from a linear operator $\LO$, we have identified an algorithm to construct a constrained branching walk model
solvable up to the $n$th moment: First, identify the weighting function
$\us(x,t)$ for which the hierarchy of moment equation closes at the
$n$th moment. To this end, solve equation
\eqref{eq:selection-function-nth-moment}, which is deterministic
nonlinear equation for the weighting function $\us(x,t)$ depending on a
space and a time variable. Second, solve the corresponding moment
equation \eqref{eq:closed-hierarchy}, which is a linear equation for
the function $\cna$ that depends on $n$ space variables
and a time variable. Once the function $\cna$ has been
obtained, any lower-order moment follows by contraction with $\un\propto\us$, as
described in Eq.~\eqref{eq:contracting}.

\section{Accounting for different subtypes}
\label{sec:fixat-prob} 

We now extend our model to account for $k$ different types of
individuals. This enables studying questions such as how does mutator
strain take over in an evolving population of bacteria even if it
confers a direct fitness detriment, or, how does a faster dispersing
mutant spreads during a growing tumor even if it might be slower
growing? Moreover, we can discuss the decay of genetic diversity and
arrive at the very important conclusion that $\us(x,t)$ is always the
fixation probability of a neutral mutation arising at position $x$ and
time $t$ by considering exchangeable subtypes.

To this end, we define the dynamics of the subtypes analogously to our original
constrained branching walk model for the total population in
Sec.~\ref{sec:derivations}: The number density of individuals of type $i$ at position $x$ at time
$t$ shall be given by a density field $c_i(x,t)$. Hence, the entire
state of the system is described by the vector $\vec
c(x,t)=\{c_i(x,t)\}_{i\in\{1,k\}}$. In each time step, a given subtype undergoes a step
of branching random walk subject to their own linear dynamics encoded
by an operator $\LO_i$ and their own fluctuations generated by a noise
field $\eta_i(x,t)$,
\begin{equation}
  \label{eq:BRW-step-subtypes}
  \tilde c_i(t+\epsilon)-c_i(t)=\epsilon \LO_i c_i + \sqrt{\epsilon\ov_i
    c_i}\,\eta_i \;.
\end{equation}
The $i$--dependence of the linear operators $\LO_i$ encode the
\emph{phenotypic} differences between types. E.g. if type $i$ would
refer to a mutator type, $\LO_i$ would include a particular mutational
operator $\mathcal M_i$ characterizing the mutator phenotype. For
instance, if mutations are modeled by diffusion, the corresponding
diffusion constant of a mutator would be larger than that of the wild
type.

The different subtypes are coupled only by the second computational
substep
\begin{equation}
  \label{eq:random_culling_subtypes}
  c_i(x,t+\epsilon)=\tilde c_i(x,t+\epsilon)(1-\lambda)\;,
\end{equation}
which ensures a global constrained defined by a weighting function
vector $\vec u=\{u_i\}_i$,
\begin{equation}
  \label{eq:constraint-subtypes}
  1=\int_x \sum_iu_i(x,t)c_i(x,t) \equiv \langle \vec u\mid \vec c \rangle\;.
\end{equation}
Notice that we have merely added another (discrete) dimension to the
problem - the type degree of freedoms. It may be checked that our
arguments to arrive at an effective stochastic differential equation
and for closing the moment hierarchy in Sec. generalize to any number
of dimensions. Thus, we can immediately restate our central
results for the extended model accounting for sub-types. 

In particular, if we choose the weighting function vector to be  $\un_i=2\us_i/\ov_i(n+1)$ with
\begin{equation}
  \label{eq:weighting-fct}
  \partial_t \us_i(x,t)=\left[\LO_i -\us_i(x,t)\right]\us_i(x,t)\;,
\end{equation}
then equation of motion for the \nth moment will be closed. If we
choose the notation
\begin{equation}
  \label{eq:moments-subtypes}
    C^{(n)}_{i_1,...,i_n}(x_1,\dots,x_n;t)\equiv\prod_{j=1}^{n}c_{i_j}(x_j,t)\;,
\end{equation}
with $i_j$ being the type of the $j^\text{th}$ number density field in
the product on the right-hand-side. The equation of motion for the
\nth moment is given by
\begin{equation}
  \label{eq:closed-hierarchy-subtypes}
  \partial_t \cna_{i_1,...,i_n}(x_1,\dots,x_n;t)=\sum^n_{j=1} \left. \left(\LO_{i_j}+\gamma(t)-\frac{2n}{n+1} \us_{i_j}
    \right)\right|_{x_j}
  \cna_{i_1,...,i_n}+\frac{2}{n+1}\sum_{j=1}^{n}\sum_{k=
    j+1}^{n}\delta_{i_j,i_k}\delta(x_j-x_k)\left \langle \vec \us\mid
    \cna_{i_1,...,i_n} \right \rangle_{k} \;.
\end{equation}
Notice that the correlations indicated by the last term only arise for
a subpopulation with itself, $i_j = i_k$.

\subsection{The neutral case and interpretation of the weighting function}
\label{sec:neutral-case}
Now, let us focus on the special case where types follow
the same dynamics in the statistical sense, 
\begin{equation}
  \label{eq:exchangeable-case}
  \LO_i=\LO\;, \qquad \us_i(x,t)=\us(x,t)\;,\qquad b_i=b\;.
\end{equation}
For such ``exchangeable'' subtypes, it is easy to see that the equation of motion
Eq.~\eqref{eq:closed-hierarchy} for the \nth moment of the total
population, $c(x,t)=\sum_ic_i(x,t)$,
is obtained upon summing left and right-hand side of
Eq.~\eqref{eq:closed-hierarchy-subtypes} over all type indices, i.e. by
carrying out $\sum_{i_1} \sum_{i_2} ... \sum_{i_n}$.

We can single out one particular subpopulation, say $i_1=\ell$ ($\ell$
for labeled), by summing the equation of motion over all other type
indices, $\sum_{i_2} ... \sum_{i_n}$. This yields
\begin{eqnarray}
  \label{eq:closed-hierarchy-1stsubtypes-rest-summed-up}
  \partial_t \overline{c_{\ell}(x_1;t)C^{(n-1)}(x_2,\dots,x_n;t)}&=&\sum^n_{j=1} \left. \left(\LO+\gamma(t)-\frac{2n}{n+1} \us
    \right)\right|_{x_j}
  \overline{c_{\ell}(x_1;t)C^{(n-1)}(x_2,\dots,x_n;t)}\\
  &+&\frac{2}{n+1}\sum_{j=1}^{n}\sum_{k=
    j+1}^{n}\delta(x_j-x_k)\left \langle  \us\mid
    \overline{c_{\ell}(x_1;t)C^{(n-1)}(x_2,\dots,x_n;t)} \right \rangle_{k} \;.
\end{eqnarray}
Note that the correlation function $\overline{c_{\ell}C^{(n-1)}}$
  satisfies the same linear equation as $\cna$ does
    Eq.~\eqref{eq:closed-hierarchy}, which we abbreviate as
\begin{eqnarray}
  \label{eq:same-eqn}
  \partial_t
  \overline{c_{\ell}(x_1;t)C^{(n-1)}(x_2,\dots,x_n;t)}&=&\mathcal G
  \overline{c_{\ell}C^{(n-1)}} \\
  \partial_t \cna(x_1,\dots,x_n;t)&=&\mathcal G
  \cna \;. \label{eq:EOM-totpop-short}
\end{eqnarray}
defining a linear (integro-differential) operator $\mathcal G$. Imagine solving for the left
eigenvector of $\mathcal G$ corresponding to eigenvalue 0,
\begin{equation}
  \label{eq:right-eigenvec}
  0=\mathcal G^\dagger M^{(n)}
\end{equation}
where $M^{(n)}(x_1,...,x_n)$ is a function of $n$ variables just like
$C^{(n)}$. Then, contracting Eq.~\eqref{eq:closed-hierarchy-1stsubtypes-rest-summed-up} with this new function
one obtains
\begin{equation}
  \label{eq:martingale-arg}
  \langle M^{(n)}|\overline{\cel C^{(n-1)}} \rangle=\text{const.}=\pf  \langle
  M^{(n)}|\cna \rangle \;.
\end{equation}
The second equality is a key step. It holds because 
$\cel C^{(n-1)}\to C^{(n)}$ on long times if fixation occurs and $\cel
C^{(n-1)}\to 0$, otherwise.

Hence, the fixation probability $\pfl$ of the labeled
subpopulation with initial density $\celn(x)$ at time $0$ is
given by
\begin{equation}
  \label{eq:pfix}
  \pfl=\frac{\langle
  M^{(n)}|\celcn \rangle}{\langle
  M^{(n)}|\cna \rangle} \;.
\end{equation}
Fortunately, the left eigenvector of $\mathcal G$  is easily constructed:
If we fully contract Eq.~\eqref{eq:EOM-totpop-short} with $n$ factors of
the weighting function $\un(x_i)$, we have to get $0$ on the LHS because of
the constraint Eq.~\eqref{eq:generalized-constraint}. Hence, the sought-after left eigenvector
can be written as
\begin{equation}
  \label{eq:left-eigenvector}
  \bigl\langle
  M^{(n)}|\cna \bigr\rangle=\left(\prod_i \int_{x_i}
    \un(x_i) \right)
  \cna(x_1,\dots,x_n) \;,
\end{equation}
Since this eigenvectors contracts to 1 with the total population, Eq.~\eqref{eq:pfix} becomes
\begin{equation}
  \label{eq:fix-probability-final}
  \pfl= \bigl \langle \un  | \celn\bigr\rangle= \frac{2\int dx \,\us(x) \celn(x)}{b(n+1)}   \;,
\end{equation}
Thus, as announced in Section \ref{sec:main-results}, a single labeled mutant at a
certain location $x$ has probability $\un(x)$ that its descendants
will take over the population on long times.

\subsection{Decay of heterozygosity}
\label{sec:decay-heterozygosity}
The diversity of labels in a population will inevitably decline,
because of the rise and ultimate fixation of one of the labels
initially present. One can capture the gradual loss of genetic
diversity by studying the expectation of the product of density fields
that correspond to different types. For instance, if there are just two types, $i\in\{1,2\}$,
the correlation function
$\cna[2]_{12}(x,y;t)=\overline{c_1(x,t)c_2(y,t)}$ satisfies
\begin{equation}
  \label{eq:homoz}
  \partial_t \cna[2]_{12}(x,y;t)=\sum^{n=2}_{j=1} \left. \left(\LO+\gamma(t)-\frac{2n}{n+1} \us
    \right)\right|_{j}
  \cna[2]_{12}\;.
\end{equation}
Notice that the right-hand side of Eq.~\eqref{eq:homoz} misses the
positive $\delta$--function term of Eq.~\eqref{eq:closed-hierarchy},
which characterized the \nth moment of the total population
density. Assuming that $\cna[2]$ (as well as $\cna[2]_{1,1}$ and
$\cna[2]_{2,2}$) have a stable stationary solution, we can conclude
that $\cna[2]_{1,2}$ will decay to $0$ at long times because of the
lacking source term. This is to be expected because the gradual
fixation of one of the two types implies that $\cna[2]_{1,2}$ has to
approach $0$ on long times. Discerning relaxation times of the
time-evolution \eqref{eq:homoz} is related to a key question in
population genetics: How fast do lineages of two individuals coalesce?

The function $C^{(2)}_{1,2}$ is closely related to the so-called heterozygosity
in population genetics. The heterozygosity in the population is the
probability that two randomly chosen individuals are of different
type, $H(t)=\int_x\int_y \overline{C^{2}_{1,2}(x,y;t)/N^2(t)}$. For our purposes, it is
much more convenient and natural to consider a variant of that quantity, 
\begin{equation}
  \label{eq:heterozygosity-variant}
  \Ht(t)=\int_x\int_y 
  u^{(n)}(x,t)u^{(n)}(y,t)C^{2}_{1,2}(x,y;t)=\langle
  u^{(n)}|c_1\rangle \langle
  u^{(n)}|c_2\rangle=p(t)q(t)
\end{equation}
with $p(t)\equiv \langle u^{(n)}|c_1\rangle$ and $q(t)\equiv \langle
u^{(n)}|c_2\rangle$. These quantities have nice properties. At any
time, both $p(t)$ and $q(t)$ represent the probability of fixation of
the respective subpopulations. Accordingly, we have $p(t)+q(t)=1$,
ensured by the global constraint. Thus, $\Ht(t)$ is similar to a
heterozygosity in neutral populations and, under certain conditions
discussed below, even converges against the actual heterozygosity.

An equation of motion for the expectation $\overline{\Ht(t)}$ can be derived by contracting
Eq.~\eqref{eq:homoz} twice with $\un$ and using the
equation of motion of $\un$, Eq.~\eqref{eq:selection-function-nth-moment-mr},
\begin{equation}
  \label{eq:heterozygosity-var-eom}
  \partial_t \overline{\Ht(t)}=-\frac{n-1}{(n+1)^2}\left(\overline{p(t) \left\langle \us^2|c_2\right\rangle}+\overline{q(t)\left\langle \us^2|c_1\right\rangle}\right)\;.
\end{equation}
Notice that the right-hand side is negative always. Thus, again, we see
that the heterozygosity will necessarily decay.

\subsubsection{Separation ansatz}
\label{sec:sep-ansatz}
As in Eq.~\eqref{eq:homoz} for $n=2$, one can easily see that the equation of
motion for $\cna_{i_1,\cdots,i_n}$ is separable for the $n$
components if $i_j\neq i_k$. It therefore admits a solution of the
simple form 
\begin{equation}
  \label{eq:sep-ansatz}
  C^{(n)}_{i_1,\cdots,i_n}=\prod_{k=1}^n f_{i_k}(x_{i_k};t)
\end{equation}
with $f_{i_k}(x_{i_k};t)$ satisfying a one-dimensional linear equation
\begin{equation}
  \label{eq:sep-ansatz:dynamics}
  \partial_t f_{i_k}(x;t)=\left(\LO+\gamma(t)-\frac{2n}{n+1} \us
    \right) f_{i_k}(x;t)
\end{equation}
subject to the initial conditions $f_{i_k}(x;0)=c_{i_k}(x;0)$.
By contracting with $\us$ and using its dynamics, $-\partial_t\us = \bigl(\LO^\dagger-\us\bigr)\us$, 
\begin{equation}
  \label{eq:sep-ansatz-lyapu}
  \partial_t \langle \us|f_{i_k}\rangle=-\frac{n-1}{n+1} \langle \us^2|
  f_{i_k}\rangle \;,
\end{equation}
we see that $f(x;t)$ must be continuously decaying with time. 

On long times, we can assume that $f(x,t)\sim a(t) \psi(x)$, where
$\psi(x)$ is the eigenfunction to the largest eigenvalue of Eq.~\eqref{eq:sep-ansatz:dynamics}. Inserting
this asymptotic behavior into Eq.~\eqref{eq:sep-ansatz-lyapu} yields an
exponential decay $\partial_t a=- a/\tau_c$ of the mode amplitude
$a(t)$ with a decay time given by
\begin{equation}
  \label{eq:longest-relax}
  \tau_c=\frac{n+1}{n-1}\frac{\langle w|\psi\rangle}{\langle w^2|\psi\rangle} \;,
\end{equation}
Intuitively, this decay time describes how long it takes until
significant fraction of the population has coalesced. Thus, one
expects $\tau_c$ to depend on the fundamental parameters of the
considered model in just the same way as the population coalescence
time. The numerical coefficient, of course, will be different by a
factor of order 1.
 
Due to the one-dimensional nature of Eq.~\eqref{eq:sep-ansatz}, it is
possible to obtain good approximations to the longest relaxation time
for various models, either by directly solving the eigenvalue problem
or by guessing the function $\psi$ in Eq.~\eqref{eq:longest-relax}. We
will provide a heuristic calculation for the case of a diffusive
kernel in the limit of large populations (or fast waves).

\subsubsection{Time-scale separation}
\label{sec:time-scale-separ}
In most models of noisy traveling waves, one has found empirically a
time scale separation: For large $\ln N$, the longest relaxation time
of the mean density field is much shorter than the coalescence
time. In the case of adaptation waves and invasion waves with
diffusion kernel, this is evident from gap in the two lowest
relaxation times in our numerical results in Fig.~\ref{fig:decaytimes}, \ref{fig:fisherwaves}B (also
see SI Fig.~\ref{fig:spectrum-SI}).

In the presence of such a time-scale separation, we can approximate $c_{1}(x,t)\approx p(t) \overline{c}(x,t)$ and
$c_{2}(x,t)\approx q(t)\overline{c}(x,t)$. For the purpose of using
these approximations in the terms involving $\langle \us^2|c_i\rangle$
in Eq.~\ref{eq:heterozygosity-var-eom}, we need them to be good in the high fitness
tail. Then, we obtain
\begin{equation}
  \label{eq:heterozygosity-var-eom:approx}
  \partial_t \overline{\Ht(t)}\approx -2\frac{n-1}{(n+1)^2}\overline{p(t)q(t)} \left\langle \us^2|\overline{c}\right\rangle\;,
\end{equation}
suggesting that the time scale for coalescence scales as 
\begin{equation}
  \label{eq:heterozygosity-decay-time}
  \tau_{0,\,\mathrm{approx}}\propto \left\langle \us^2|\overline{c}\right\rangle^{-1}\;.
\end{equation}
This approximation indeed seems to approach the correct time scale for
large $\ln N$, as can be appreciated from Figs. \ref{fig:decaytimes} and
\ref{fig:fisherwaves}B.

\section{Discussion}
\label{sec:discussion}
The ecological and evolutionary fate of populations often depends on a
small number of ``pioneers'', distinguished by their growth rates,
migration rates, location, or other characteristics correlated with
long-term survival.  Most analyses of these inherently stochastic problems
have focussed on the mean behavior of the population, which
sensitively depends on fluctuations in the pioneer populations. Yet,
the mean behavior says little about any given realization, the
variability between realizations and their correlation times.

Here, we have shown that fluctuations can be analyzed in principle, if
one relies on minimal models that reduce the dynamics to two essential
ingredients: (1) Birth, death and jumps give rise, effectively, to a
branching random walk. (2) A non-linear population regulation makes
sure that those branching processes do not get out of control. For
such constrained branching random walks, we have provided a general
route towards analyzing fluctuations. The basic idea of the method is
to adjust the population control, an essential non-linearity, in such
a way that the equations describing correlation functions of order $n$
are closed.

Our method can be used to elucidate variability between replicates in
evolution experiments as well as the genetic diversity within a
population. To provide specific results, we have focussed on simple
models of adaptation and of invasions. In both cases, we have found
that the decay of genetic diversity scales as a power of the logarithm
of the population size for large population sizes. Higher moments show
a marked anticorrelation between the dynamics in the tip and the bulk
of the wave. Moreover, we found that, for the models analyzed, the
time scale for the decay of higher order correlations, such as the
genetic heterozygosity, is much longer than the time the population
wave needs to equilibrate at a given speed or population size. The
presence of such a time scale separation simplifies the analysis of
coalescence times considerably.

The ensemble of the resulting tuned models is complementary to
established models of adaptation. While the latter have fixed
population and fluctuating speeds of adaptation, the former has a
fluctuating population size but fixed wave speeds. The resemblance of
both ensembles relies on the fact that the fluctuations occur on time
scales long compared to the relaxation time of the population
wave. Quantities that only depend on the mean logarithm of the
population size, such as the wave speed or the coalescence time, thus
agree asymptotically in both ensembles, see
e.g.~Fig.~\ref{fig:popsizespeed}, \ref{fig:fisherwaves}.

One might wonder about the net-effect of noise on models of adaptation
and other traveling waves. If one is only concerned with the mean,
many previous works have assumed that the effect of noise can be
summarized by an effective cutoff in the tip of the
wave~\cite{tsimring1996rna,brunet1997shift,cohen2005front}. This
cutoff effect can be explicitly seen in the closed first moment
equation of tuned models, as was already pointed out in
Ref.~\cite{hallatschek2011noisy}. However, what is the effect of noise
on higher-order correlations? Our general formulation in
Eq.~\eqref{eq:closed-hierarchy} of the \nth moment exhibits, in
general, two terms with different signs that are unexpected in a
deterministic framework. One term tends to generate positive
correlations between nearby individuals (in fitness space). These
correlations then dissipate over distances due to the term with
negative sign. Importantly, the correlation term becomes dominating in
the tip of the wave due to its density dependence.  The net-effect of
fluctuations on the correlations emphasizes the complex nature of
fluctuations, which only to the lowest order can be captured by a
simple cutoff term in an effective Liouville operator.

The branching random walk contains a parameter $b$, the variance in
offspring numbers per generation, that effectively measures the
strength of genetic drift.  Surprisingly, the noise-induced terms do
not depend on this parameter $b$. This means that the noise-induced
terms are not small, in general, even if the parameter $b$ is small,
such that a controlled small-noise perturbation analysis is not
possible. This reinforces the observation that noise is a singular
perturbation that fundamentally impacts the outcome of ecological and
evolutionary processes.

Our method of model tuning is quite versatile as it applies to any branching random
walk subject to a global constraint. This includes models that combine
ecology and evolution~\cite{barton2013modelling,pelletier2009eco} or
epistatic models in which mutations are not simply additive but might
interact~\cite{tenaillon2014utility}. However, for more complex
scenarios of interest to evolutionary biologists, one would like to
introduce additional non-linearities. For instance, sex and
recombination is a quadratic non-linearity as it depends on the the
probability density of two different individuals finding each other
and mating~\cite{neher2010rate}. In evolutionary game theory, one is
interested in mutants that have a frequency-dependent
advantage~\cite{reiter2014range,kussell2014non}. The fitness of producers of a common
good depends on the frequency of producers. This, again, introduces a
non-linearity, which is quadratic in the simplest case.

Such non-linearities cannot be included in an exact way because they
generate higher-order terms. However, it may be a useful approach to
build them in and truncate the moment hierarchy at an appropriate
order provided one can show that the neglected terms really are small.
We believe that such reasoning should work, typically, if the
non-linearities do not strongly influence the dynamics in the small
density regions where the noise strength is large. A truncation scheme would,
in this case, amount to matching a stochastic but linear description
of the wave tip with a deterministic but non-linear bulk of the wave.
We would welcome future work to examine these possibilities.

\section{Acknowledgments}
\label{sec:acks}
Thanks to Peter Pfaffelhuber for useful discussions and making us
aware of Ref.~\cite{katzenberger1991solutions}. This work was
partially supported by a Simons Investigator award from the Simons
Foundation (O.H.), the Deutsche Forschungsgemeinschaft via Grant HA
5163/2-1 (O.H.).



\appendix

\begin{figure}[tb]
\begin{center}
\includegraphics[width=\textwidth]{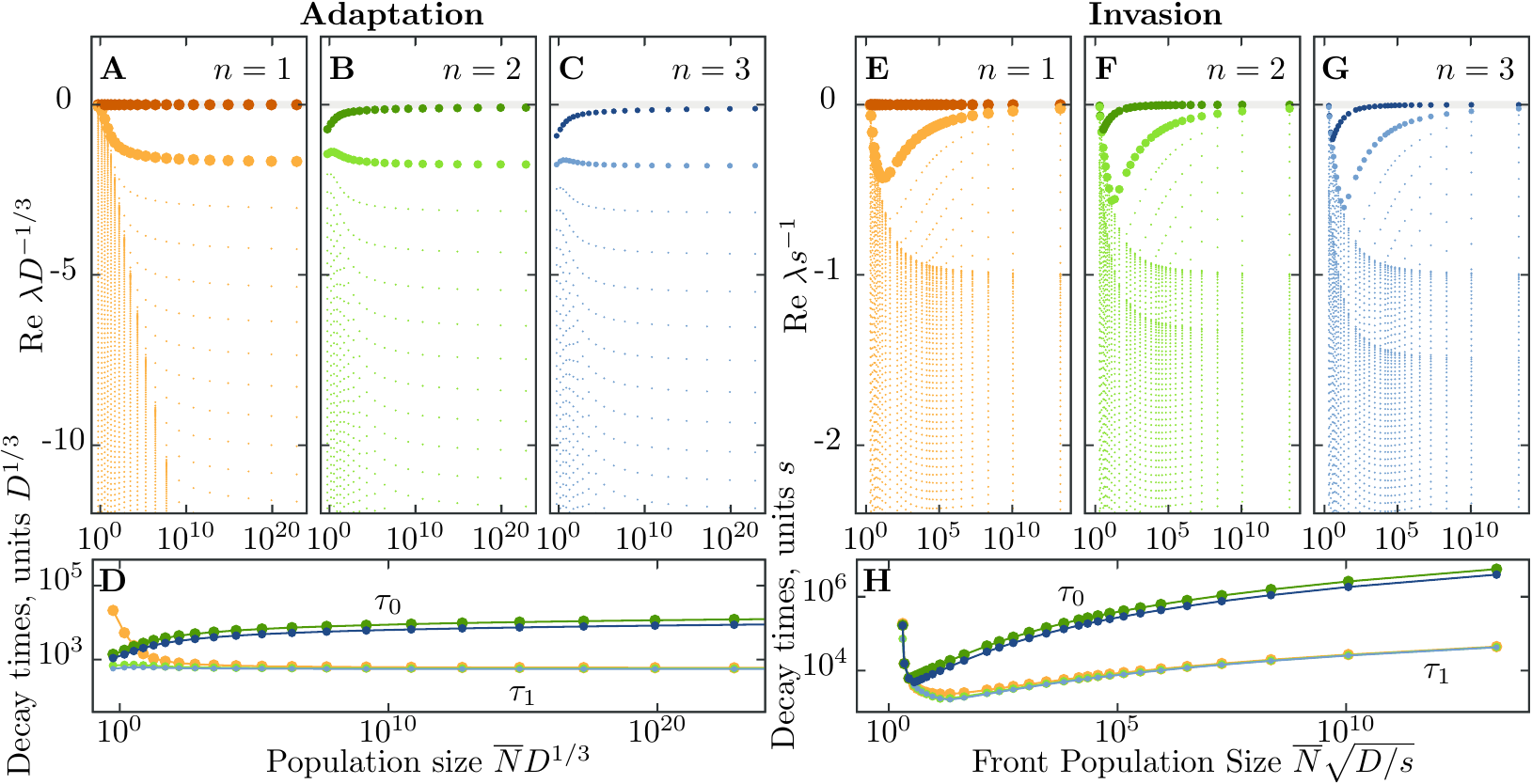}
\caption{\label{fig:spectrum-SI} \textbf{Relaxation rate spectra for
    models of adaptation and invasion.} This figure quantifies the
  relaxation rates of correlations among labeled subpopulations in
  models of adaptation (left) and invasion (right). The relaxation
  rates $\lambda_i$ for the \nth model were found from a spectral
  analysis of the equation of motion for $n$ separable subpopulations in the \nth correlation
  function, cf. Eq.~\eqref{eq:sep-ansatz:dynamics}. All eigenvalues are real and negative, i.e., they lead to
  the decay of correlations. Each relaxation rate $\lambda_i$
  corresponds to a relaxation time $\tau_i=1/\lambda_i$.  {\bf D} and
  {\bf F} show the behavior of the two slowest decay rates $\tau_0$
  and $\tau_1$ together with an approximation for $\tau_0$. The ratio
  $\tau_0/\tau_1$ controls the time scale separation between
  coalescence (slow) and wave profile relaxation (fast). 
  In addition, the effects of chosing a particular $n$ for the closure of the moment hierarchy vanishes for large population sizes (or wave speeds).
  Relaxation time scales of eigenmodes become more and more similar.
  Figs.~\ref{fig:decaytimes} and \ref{fig:fisherwaves} depict only the first two timescales $\tau_0$ and $\tau_1$ (here shown as bold dots in the upper panels and emphasized again in lower panels) and concentrate on the case $n=2$.
  }
\end{center}
\end{figure}

\section{Explicit equations of motion in tuned models}
\label{appendix:explicit}
In the main text, we presented general equations for the correlation functions $\cna$ and the weighting function $\us$.
Numerical results have been computed for the two cases $n=1$ and $n=2$.
For reference, here we state the explicit equations of motion for
both, adaptive and invasive, waves.

\subsection{Adaptation waves}
The first step is always to calculate the weighting function $\us$ in a comoving frame with speed $v$:
\begin{equation}
\label{eq:app:explicit:adapt:weigthing}
0 = -v\partial_x\us(x) + D\partial_x^2\us(x) + x\us(x)-2\us(x)^2\;.
\end{equation}
Here, the term with multiplication of fitness, $x\us(x)$, indicates the selection term in adaptive waves, which has to replaced with $s\Theta(x)\us(x)$ for invasive waves, see below.

\subsubsection{Model tuned to the first moment ($n=1$)}
After obtaining the general form of the weighting function $\us$, the mean stationary population density follows as
\begin{equation}
\label{eq:app:explicit:adapt:1st}
0 = v\partial_x\cna[1](x) + D\partial_x^2\cna[1](x) + x\cna[1](x) -\us(x)\cna[1](x)\;,
\end{equation}
a result that has already been described in Ref.~\cite{hallatschek2011noisy}.

\subsubsection{Model tuned to the second moment ($n=2$)}
If we choose $n=2$, the model is tuned to have a closed 2-point correlation function $\cna[2](x,y)$, which is governed by
\begin{eqnarray}
\nonumber 0 &=& v\partial_x\cna[2](x,y) + v\partial_y\cna[2](x,y) + D\partial_x^2\cna[2](x,y) + D\partial_y^2\cna[2](x,y) \\
\label{eq:app:explicit:adapt:2nd} && + \bigl(x+y-4\us(x)/3-4\us(y)/3\bigr)\cna[2](x,y) + 2/3\,\delta(x-y)\int dz~\us(z)\cna[2](x,z)
\end{eqnarray}
In this case, the mean stationary population density is obtained by
contraction, $\cna[1] = \bigl\langle
u^{(2)}\mid\cna[2]\bigr\rangle$.

\subsection{Invasion waves}
For invasion waves the weighting function $\us$ is the solution to
\begin{equation}
\label{eq:app:explicit:inv:weighting}
0 = -v\partial_x\us(x) + D\partial_x^2\us(x) + s\Theta(x)\us(x) -\us(x)^2\;,
\end{equation}
which again sets the speed $v$ of the comoving frame as parameter.
In addition, tuning the model for the first moment, $n=1$, yields the equation of motion for the stationary mean population density $\cna[1](x)$,
\begin{equation}
\label{eq:app:explicit:inv:1st}
0 = v\partial_x\cna[1](x) + D\partial_x^2\cna[1](x) + s\Theta(x)\cna[1](x) - \us(x)\cna[1](x)\;.
\end{equation}
Choosing $n=2$ for invasion waves leads to
\begin{eqnarray}
\nonumber 0 &=& v\partial_x\cna[2](x,y) + v\partial_y\cna[2](x,y) + D\partial_x^2\cna[2](x,y) + D\partial_y^2\cna[2](x,y) \\
\label{eq:app:explicit:inv:2nd} && + \bigl(s\Theta(x)+s\Theta(y)-4\us(x)/3-4\us(y)/3\bigr)\cna[2](x,y) + 2/3\,\delta(x-y)\int dz~\us(z)\cna[2](x,z)\;.
\end{eqnarray}

\section{Numerical methods}
\label{appendix:numerics}

Although asymptotic analyses of noisy traveling wave models are often
possible, closed form solution of either the fixation
probability $\us$ or correlation function $\cna$ are usually out of
reach.  Numerical methods can help alleviating this problem, as with
tuned models we are able to state at least \emph{exact} moment equations,
which do not need an further approximations or assumptions.

In order to solve the governing equations of our tuned model
numerically, we implemented an algorithm based on a multi-dimensional
Newton-Raphson (NR) iteration scheme
\cite{press2007numerical,geyrhofer2014quantifying}.  Here, we recount
the basic steps of this scheme.

For the strictly one-dimensional case, $n=1$, the numerical solution
involves the roots of a set of $M$ equations in $M$ variables,
\begin{equation}
\label{eq:numapp:eom}
f_i(y_1,\dots,y_M) = 0\;,~~~1\leq i \leq M\;,
\end{equation}
which represent the steady state equations of motion, e.g.,
Eq.~\eqref{eq:app:explicit:adapt:weigthing} and
Eq.~\eqref{eq:app:explicit:adapt:1st} for the case of adaptation and
$n=1$. The variables $y_i$ denote the value of the desired weighting
function or correlation function at lattice point $i$, respectively.

For a small deviation $\delta y_i$ in one of the variables, one can expand $f_i$ into a series,
\begin{equation}
\label{eq:NR:seriesexpansion}
f_i(y_1,\dots,y_i+\delta y_i,\dots,y_M) = f_i(y) + \partial_{y_i}f_i(y) \delta y_i + O(\delta y_i^2)\;.
\end{equation}
In the NR scheme, one iterates the current guess for the solution $y^\mathrm{old}$ to obtain $y^\mathrm{new} = y^\mathrm{old} + \delta y$.
The small difference $\delta y$ is extrapolated by assuming the new solution $y^\mathrm{new}$ fulfills the equation of motion, $f_i(y^\mathrm{new})\stackrel{!}{=}0$, while truncating \eqref{eq:NR:seriesexpansion} after the \emph{linear} term.
This leads to the equation $0 = f_i(y^\mathrm{old}) + \partial_{y_i}f_i(y^\mathrm{old})\delta y_i$.
Thus, a single step comprises of evaluating the expression
\begin{equation}
\label{eq:NR:singlestep}
y_i^\mathrm{new} = y_i^\mathrm{old} - \frac{f_i(y_{i-1}^\mathrm{old},y_{i}^\mathrm{old},y_{i+1}^\mathrm{old})}{\partial_{y_i}f_i(y_{i-1}^\mathrm{old},y_{i}^\mathrm{old},y_{i+1}^\mathrm{old}) }
\end{equation}
for each lattice point.  In order to ensure a better convergence, the
solutions $y_i^\mathrm{new}$ for even and odd indices are computed
consecutively.  In the notation of \eqref{eq:NR:singlestep} we also
made use of a simplifying fact: the diffusion approximation (for the
mutation process in adaptive wave and movement in the invasive waves)
leads only to a ``local'' coupling, such that the equations of motion
only depend on the values of $y$ at the focal lattice site $i$ and the
two neighboring ones, $f_i(y_{i-1},y_i,y_{i+1})$.  For the case of
adaptation waves, we discretize Eq.~\eqref{eq:app:explicit:adapt:weigthing}
and obtain the required expressions for the weighting function $\us$,
\begin{eqnarray}
\label{eq:NR:f} f_i\bigl(\us_{i-1},\us_i,\us_{i+1}\bigr) &=& \bigl(D/dx^2 + v/2dx\bigr)\us_{i-1} + \bigl(x_i-2D/dx^2\bigr)\us_i + \bigl(D/dx^2-v/2dx\bigr)\us_{i+1}-\us_i^2\;,\\
\label{eq:NR:fprime} \partial_{\us_i}f_i\bigl(\us_{i-1},\us_i,\us_{i+1}\bigr) &=& \bigl(x_i-2D/dx^2\bigr) - 2\us_i\;,
\end{eqnarray}
which have to be inserted into Eq.~\eqref{eq:NR:singlestep}.
For obtaining solutions, the lattice spacing $dx$ has to be chosen
small enough, that the Right-Hand-Side of \eqref{eq:NR:fprime} is negative on the whole lattice and does not change sign for any $x_i =i\,dx$.
The range of $i$ has to be adjusted to fit all characteristic features of the profiles onto the $M$ lattice points.
Similar expressions to \eqref{eq:NR:f} and \eqref{eq:NR:fprime} hold for the mean stationary population density $\cna[1]$ after discretizing \eqref{eq:app:explicit:adapt:1st}.

For the higher dimensional correlation functions $\cna$, $n>1$, the method
can be extended in a straightforward fashion.  For instance, for $n=2$
one has $M\times M$ variables $y_{ij}$ and $M\times M$ functions
$f_{ij}$.  Each dimension only adds two additional variables in the
equation of motion,
$f_{ij}(y_{i-1,j},y_{i,j-1},y_{i,j},y_{i+1,j},y_{i,j+1})$.  Note that
the discrete limit of the Dirac-delta for correlations in the last term of Eq.~\eqref{eq:app:explicit:adapt:2nd} is
given by $\delta(x_i - x_j) = 1/dx\,\delta_{ij}$.

An improvement of this algorithm, that utilizes not only the ``local''
derivative $\partial_{y_i}f_i$, but the whole Jacobian with entries
$\mathcal J_{ij} = \partial_{y_i}f_j$, is often needed for extended
mutation kernels $\mu(y)$ in \eqref{eq:mutational-operator}
\cite{geyrhofer2014quantifying,geyrhofer2015oscillations}.
In these cases (and for reasonable parameter values), Eq.~\eqref{eq:NR:fprime} changes its sign twice on any lattice, regardless of the choice of lattice spacing $dx$, which renders this ``local'' approximation \eqref{eq:NR:singlestep} unusable.
However, \eqref{eq:NR:singlestep} suffices for all present purposes.

In Figure \ref{fig:spectrum-SI} we displayed the spectral decomposition
of the linear operator governing the equation of motion for the
stationary mean population density. In this case, the governing
(discretized) equations \eqref{eq:numapp:eom} are given by the linear equation $0 = f_i(y_1,\dots,y_M) = \sum_j F_{ij}y_j$, with coefficients $F_{ij}$ obtained from discretizing Eq.~\ref{eq:sep-ansatz:dynamics}.  The eigenvalues
$\lambda_i$ are obtained by a Schur decomposition of the matrix
$F_{ij}$, which leads to an (quasi) upper triangular matrix: Along its
diagonal it has $1\times1$ blocks with real eigenvalues and $2\times2$
blocks with its complex conjugate eigenvalues.  The existence of
complex eigenvalues depends on the mutation (or migration) scheme.
For a diffusion scheme (i.e. a second derivative) in
Eq.~\eqref{eq:mutational-operator} one obtains only real eigenvalues
numerically.  The code itself is based on the already implemented
routines in the GNU Scientific Library (GSL), in particular centered around the function {\verb gsl_eigen_nonsymmv } to provide input and parse output.

The numerical code, implemented in C, is freely available from the authors.
\section{Stochastic simulations}
\label{appendix:simulations}

For the stochastic simulations of adaptation waves, the continuous density $c_t(x)$ in
fitness space is discretized into bins on a regular one-dimensional
lattice.  All individuals in the interval $\bigl[x_i,x_{i+1}\bigr]$
with $x_i = i\,dx$ are counted in the \emph{occupancy vector} $n_i$,
\begin{equation}
\label{eq:simcode:occupancy:def}
n_i = c_t(x_i)dx\;.
\end{equation}
The occupancies are updated in discrete time steps of length $\epsilon$, which encompass the dynamics in the stochastic equation \eqref{eq:BRW-genotype} and the subsequent step to limit population sizes, \eqref{eq:pop-size-fixed} or \eqref{eq:generalized-constraint}.
The action of these equations consists of three sub-steps in the algorithm, indicated by superscripts in subsequent equations.
In each of those steps the occupancies (can) change.

First, the mean (deterministic) change due to a comoving frame, mutations and selection is applied,
\begin{equation}
\label{eq:simcode:step1}
\Delta^{(1)}n_i/\epsilon = \frac{v}{2dx}\bigl(n_{i+1}-n_{i-1}\bigr)+\frac{D}{dx^2}\bigl(n_{i-1}-2n_i-n_{i+1}\bigr) + \bigl(x_i-x_0(t)\bigr) n_i\;.
\end{equation}
The first term for the comoving frame is only used in simulations with
a tuned constraint.  The next
term represents modifications in fitness due to mutations, while the
last term represents growth due to selection.  For the latter, we have
to distinguish again a fixed population size constraint and our tuned
constraint.  The offset $x_0(t)$ in the selection term is either set to
the mean fitness $x_0(t) = \sum_i x_in_i/N$ in the fixed population
size constraint, or set to $x_0(t)=0$ as we incorporated the change in
mean fitness already with the comoving frame.  For the case of
invasion waves, the selection term is replaced by
$s\Theta\bigl(x_i-x_0(t)\bigr)n_i$.  There the offset $x_0(t)$ is the
position of the front, which we define as $x_0(t) =
\sum_ix_in_in_{-i}/\sum_in_in_{-i}$ for a fixed population size
constraint.  Again, we have $x_0(t)=0$ for the tuned constraint in its
comoving frame.

In the next sub-step, the randomness due to birth and death events (i.e. genetic drift) further modifies all occupancies $n_i$,
\begin{equation}
\label{eq:simcode:step2}
\Delta^{(2)}n_i/\sqrt{\epsilon} = \sqrt{2}\bigl(\mathrm{Poisson}(n_i)-n_i\bigr)\;,
\end{equation}
where $\mathrm{Poisson}(n_i)$ is a Poisson-distributed random number
with parameter $n_i$.  This particular form of the noise (with $n_i$
already updated from mutations and selection), ensures that (i)
occupancies $n_i$ do not drop below zero, (ii) the mean value of the
noise is zero and (iii) the variance at a lattice site $i$ amounts to
$2\epsilon n_i$ per time step $\epsilon$.  While the introduction of
occupancies $n_i$ instead of a density $c_i$ is irrelevant for most of
the algorithm, it is convenient for this last feature (iii).  The
correlations of the noise \eqref{eq:eta-eta-corrs} are given by the
expression $\overline{\eta_t(i\,dx)\eta_{t'}(j\,dx)} = \delta_{tt'}
(1/dx\,\delta_{ij})$ in the discretized simulations, where the factor
$1/dx$ is then scaled already into the $n_i$.

In the last sub-step, the population is scaled uniformly to comply with its constraint,
\begin{equation}
\label{eq:simcode:step3}
\Delta^{(3)}n_i = \Bigl(\frac{1}{\sum_ju_jn_j}-1\Bigr)n_i\;.
\end{equation}
For the fixed population size constraint in section \ref{sec:second-subst-comp}, we simply set $u_j = N^{-1}$ for all $j$ (or, for invasion waves $u_j=N^{-1}$ for $j\,dx > x_0(t)$ and $u_j=0$ otherwise).
In the case of tuned models, we first solve the equation of motion for the fixation probability in a comoving frame.
To obtain this numerical solution for $\us$, we utilize the code presented in appendix \ref{appendix:numerics}.

Simulation code, written in C, is available upon request from the authors.

\section{Measuring fixation probabilities}
\label{appendix:measurefixation}
In Fig.~\ref{fig:measurefixation}{\bf B, C} we presented confidence intervals for the fixation probability, obtained via measurements of fixation and extinction events.
\textit{A priori}, counting events leads to an average value for the fixation probability, which might or might not be close to the expected (theoretical) value.
In order to compute confidence intervals, additional assumptions have to be made, explained below.

\begin{figure}
  \includegraphics[width=.5\columnwidth]{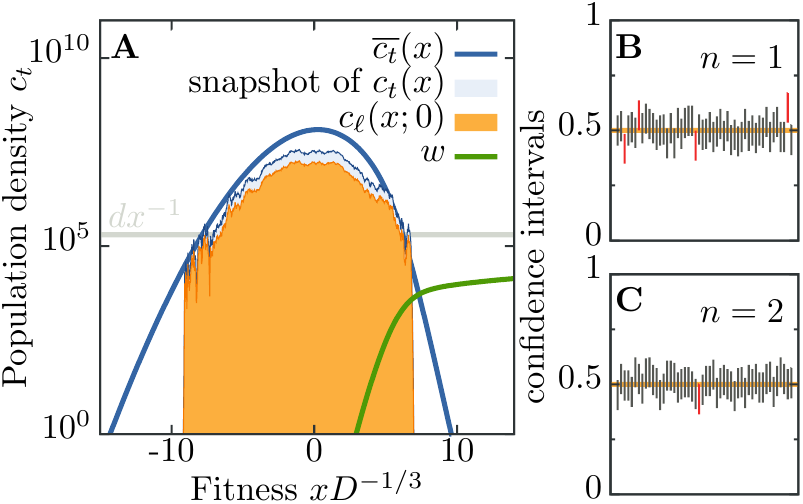}
  \caption{\label{fig:measurefixation:half}
  \textbf{Fixation and extinction measurements with the trivial labelling of $c_\ell(x,0) = c(x,0)/2$.}
  With such a labelling we obviously expect half of all simulations to end in fixation and half to end in extinction of the subpopulation $c_\ell$.
  The obtained confidence intervals serve as indicator for the noisiness of such measurements, when comparing results to the position depend labelling (cf. Eq.~\eqref{eq:measurements:posdeplabel}) presented in Fig.~\ref{fig:measurefixation}.
  In general, however, the initial hypothesis of a fixation probability of $1/2$ is corroborated by simulation results (Panels {\bf BC}).
  }
\end{figure}

After generating different starting conditions $c_t$ (snapshots from stochastic simulations), we label subpopulations $c_\ell$ in the nose of the wave, such that
\begin{equation}
\label{eq:measurements:def}
\bigl\langle \un \mid c_\ell\bigr\rangle = 1/2\;.
\end{equation}
There are, of course, many ways to label a subpopulation such that the
expected fixation probability is $50\%$. The simplest way is to label
$50\%$ of the population in each bin, which unsurprisingly yields a
fixation probabilities very close to $50\%$ (cf. Fig.~\ref{fig:measurefixation:half}). However, this naive labelling protocol does not test our
predictions for the spatial dependence of the fixation probability. To
test the accuracy of our predictions in the spatially varying region
of the fixation probability, we label the population in the tip of the
wave, using the following form
\begin{equation}
\label{eq:measurements:posdeplabel}
c_\ell(x;t) = \frac{c(x;t)}{1+\exp\bigl(-(x-x_\ell)/\delta\bigr)}\;.
\end{equation}
Here, $x_\ell$ determines the position of the labelling and $\delta$
its steepness. To avoid artifacts associated with the discreteness, we
choose the length scale $\delta$ such that on
the order of $10$ bins, typically, contain both, labelled and
unlabelled, subpopulations.  The crossover $x_\ell$ is iteratively
adjusted until the condition in Eq.~\eqref{eq:measurements:def} is met
with sufficient accuracy (usually, we demand a value close to machine
precision, $10^{-10}$).

From this configuration, we run the stochastic time evolution of the
population until the labelled subpopulation either reaches fixation or
goes extinct.  In accordance with our interpretation of $\un$ as
fixation probability, we expect the labeled population to fix in half
of the simulation runs and to go extinct otherwise.  For definiteness,
we abort simulations when one of the thresholds $\bigl\langle \un\mid
c_\ell\bigr\rangle=1-10^{-4}$ or $\bigl\langle\un\mid
c_\ell\bigr\rangle = 10^{-4}$ is exceeded.  We count such events as
fixation and extinction, respectively.

After having amassed such simulation evidence, we can use Bayesian
inference to check if our assumption of the interpretation of $\un$ as
fixation probability is consistent \cite{ohagan1994advanced}.  The
distribution of the fixation probability $\pfl$ of the sub-population,
given the (simulation-) data is computed as
\begin{equation}
  \mathbb P\bigl[\,\pfl\,\vert\, \mathrm{data}\,\bigr] \sim \mathbb P\bigl[\,\mathrm{data}\,\vert\,\pfl\,\bigr]\,\mathbb P\bigl[\,\pfl\,\bigr]\;,
\end{equation}
using Bayes' theorem. Here, the likelihood $\mathbb
P\bigl[\,\mathrm{data}\,\vert\,\pfl\,\bigr]$ of observing either an
extinction or fixation event is a simple binomial distribution: when
having $N$ trials with $X$ fixation events, the likelihood is $\mathbb
P\bigl[\,X\,\mathrm{ Fixation\,events}\vert\, \pfl\,\bigr]=
\binom{N}{X} (\pfl)^X(1-\pfl)^{N-X}$.  Furthermore, we assume a flat prior $\mathbb
P\bigl[\pfl\bigr]$ for the fixation probability $\pfl$, ignoring any
knowledge about its value at the beginning.  Such a flat prior can
also be cast as a Beta distribution (incidentally a conjugate prior
\cite{ohagan1994advanced}), which in its general form is given by
\begin{equation}
\mbox{Beta}\bigl(Y;\alpha,\beta) \propto \frac{\Gamma(\alpha+\beta)}{\Gamma(\alpha)\Gamma(\beta)}Y^{\alpha-1}(1-Y)^{\beta-1}\;.
\end{equation}
Using the two hyperparameters $\alpha
=1$ and $\beta=1$ we arrive at the uniform (flat) distribution on
$\bigl[0;1\bigr]$.  Thus, the posterior distribution for the fixation
probability $\pfl$ of the sub-population can also be written as
Beta-distribution:
\begin{equation}
\label{eq:fixprob_posterior}
\mathbb P\bigl[\,\pfl\,\vert\, X \,\mathrm{ Fixation\,events}\,\bigr]\sim (\pfl)^{X+\alpha-1}(1-\pfl)^{N-X+\beta-1}\;,~~\alpha=\beta=1\;,
\end{equation}
up to a normalization factor. From this posterior \eqref{eq:fixprob_posterior} we can evaluate the $95\%$ confidence interval, and check if our assumption $\pfl=1/2$ is within its range.
Increasing the value of $\alpha=\beta > 1$ would increase the
certainness of our initial hypothesis, $\pfl=1/2$, that we put into
the model, narrowing the distribution
\eqref{eq:fixprob_posterior}.

\end{document}